\documentclass[12pt]{article}
\usepackage{comment}
\usepackage{amsmath}
\usepackage{amssymb}
\usepackage{graphicx}
\usepackage{here}
\usepackage{subcaption}
\usepackage{url}
\usepackage[sort&compress, numbers, merge]{natbib}
\usepackage{braket}
\usepackage{physics}% Added by SK
\usepackage[compat=1.1.0]{tikz-feynhand}
\usepackage{slashed}
\usepackage{mathtools}
\usepackage{booktabs}
\usepackage{standalone}
\usepackage{todonotes}

\numberwithin{equation}{section}

\usepackage[colorlinks=true, linkcolor=blue, citecolor=blue,
urlcolor=black]{hyperref} 

\begin{document}
\def\ps{\mathbf{p}}
\def\PS{\mathbf{P}}
\baselineskip 0.6cm
\def\simgt{\mathrel{\lower2.5pt\vbox{\lineskip=0pt\baselineskip=0pt
           \hbox{$>$}\hbox{$\sim$}}}}
\def\simlt{\mathrel{\lower2.5pt\vbox{\lineskip=0pt\baselineskip=0pt
           \hbox{$<$}\hbox{$\sim$}}}}
\def\simprop{\mathrel{\lower3.0pt\vbox{\lineskip=1.0pt\baselineskip=0pt
             \hbox{$\propto$}\hbox{$\sim$}}}}
\def\tr{\mathop{\rm tr}}
\def\SU{\mathop{\rm SU}}

\begin{titlepage}

\begin{flushright}
IPMU25-0049
\end{flushright}

\vskip 1.1cm

\begin{center}

{\Large \bf 
Flavor Symmetry and Proton Decay in PeV-Scale Supersymmetry
}

\vskip 1.2cm
Akifumi Chitose$^{a}$, 
Masahiro Ibe$^{a,b}$, 
and Satoshi Shirai$^{b}$
\vskip 0.5cm

{\it

$^a$ {ICRR, The University of Tokyo, Kashiwa, Chiba 277-8582, Japan}

$^b$ {Kavli Institute for the Physics and Mathematics of the Universe
(WPI), \\The University of Tokyo Institutes for Advanced Study, \\ The
University of Tokyo, Kashiwa 277-8583, Japan}

}

\vskip 1.0cm

\abstract{
Supersymmetry beyond the TeV scale offers several theoretical and phenomenological advantages, such as accommodating the observed Higgs mass and alleviating the flavor and CP problems.
However, flavor and CP observables still impose stringent constraints even at the PeV scale, motivating a systematic study of flavor symmetries in this regime.
In this work, we investigate nucleon decay induced by dimension-five operators in supersymmetric standard models and examine how flavor symmetries, particularly of the Froggatt-Nielsen type, can suppress these operators.
We perform a Bayesian analysis combining flavor, CP, and proton-decay observables to quantify the allowed parameter space and identify characteristic predictions.
Our results demonstrate that a multi-messenger approach, integrating flavor, CP, and baryon-number-violating observables, is essential for probing the underlying structure of supersymmetry beyond the TeV scale.
}

\end{center}
\end{titlepage}

\tableofcontents
%-----------------------------------------------------
\section{Introduction}  
\label{intro}
%-----------------------------------------------------

Supersymmetric extensions of the Standard Model (SSM) have long been regarded as one of the most compelling candidates for physics beyond the Standard Model (SM), primarily due to their ability to address the naturalness problem associated with the weak scale~\cite{Maiani:1979gif,Dimopoulos:1981zb,Sakai:1981gr}. 
This led to widespread attention on weak-scale supersymmetry as a promising framework.

However, it was already recognized, even before the LHC era, that weak-scale supersymmetry faces several theoretical challenges.
SSM generally induces flavor-changing neutral currents (FCNCs) and additional sources of CP violation, which impose severe constraints on weak-scale supersymmetry (see e.g., Ref.\,\cite{Gabbiani:1996hi}).
Another long-standing issue arises from proton decay: even with $R$-parity conservation, dimension-five proton decay operators can lead to excessively rapid decay rates~\cite{Sakai:1981pk,Weinberg:1981wj}.

Furthermore, the presence of a weak-scale gravitino gives rise to serious cosmological problems, commonly referred to as the gravitino problem.
It had also been known that reproducing the observed dark-matter relic abundance with a thermal Wino or Higgsino requires a soft supersymmetry-breaking mass scale above the TeV range.
From these perspectives, the possibility of supersymmetric models beyond the TeV scale had already been actively explored well before the advent of LHC experiments~\cite{Wells:2003tf,Arkani-Hamed:2004ymt,Giudice:2004tc,Arkani-Hamed:2004zhs,Wells:2004di}.

The discovery of the 125 GeV Higgs boson further accelerated the exploration of supersymmetric models beyond the TeV scale, including their implications for grand unification, supersymmetry-breaking sectors, and cosmology.
In particular, supersymmetric frameworks based on the anomaly-mediated supersymmetry breaking (AMSB) mechanism~\cite{Randall:1998uk,Giudice:1998xp}, which typically predict PeV-scale gravitino and sfermion masses, have attracted significant attention~\cite{Hall:2011jd,Ibe:2011aa,Ibe:2012hu,Arvanitaki:2012ps,Hall:2012zp,ArkaniHamed:2012gw,Nomura:2014asa}.
Such scenarios naturally alleviate the FCNC and CP-violation problems~\cite{Altmannshofer:2013lfa,McKeen:2013dma}, avoid the cosmological gravitino problem, and simultaneously accommodate the observed Higgs boson mass~\cite{Okada:1990vk,Ellis:1990nz,Haber:1990aw,Okada:1991jc,Ellis:1991zd}.
Moreover, they predict that the gaugino can be the lightest supersymmetric particle (LSP), which, under $R$-parity conservation, provides an excellent dark-matter candidate.
These features together make AMSB-based frameworks one of the most consistent and realistic realizations of supersymmetry in the current era.

Nevertheless, even in supersymmetric models beyond the TeV scale, proton decay induced by dimension-five operators remains an unresolved issue~\cite{Dine:2013nga}.
At the same time, the so-called flavor puzzle, i.e., the question of the origin of the observed flavor structure and fermion mass hierarchy in the SM, also remains open.
Addressing this problem generally requires the introduction of a flavor model, particularly one based on underlying symmetries.
These two issues are intimately connected: a suitable flavor symmetry can simultaneously account for the fermion mass hierarchy and suppress flavor-changing, CP-violating, and baryon~number-violating operators.
In particular, 
the dimension-five proton decay operators arise through the exchange of color-triplet Higgs multiplets in supersymmetric grand unified theories (GUTs), or more generally as effective interactions suppressed by a high cutoff scale, 
such as the Planck scale.
Without additional symmetries, these operators would predict proton lifetimes far shorter than current experimental bounds, highlighting the crucial role of flavor symmetries in ensuring phenomenological viability~\cite{Murayama:1994tc,Ben-Hamo:1994dha,Carone:1995xw,Carone:1996nd,Nelson:1997bt,Kakizaki:2002hs,Csaki:2013we,Nagata:2013sba}.

In this work, we perform a Bayesian analysis of the interplay between flavor symmetries and the constraints arising from dimension-five proton decay operators.
In particular, we demonstrate how the predicted proton decay signals vary across different flavor models, assuming that the dimension-five operators are not necessarily suppressed by the GUT scale but by a higher energy scale such as the Planck scale.

The remainder of this paper is organized as follows.
In Section~\ref{sec:setup}, we describe the setup of the SSM under consideration, including the Yukawa interactions, soft supersymmetry-breaking terms, and dimension-five operators relevant to proton decay.
Section~\ref{sec:proton_flavor} discusses the current experimental limits from proton decay searches and FCNC/CP constraints, 
and outlines our Bayesian framework for evaluating their combined implications.
In Section~\ref{sec:FN}, we introduce the Froggatt-Nielsen (FN) mechanism and examine its impact on sfermion masses, baryon~number-violating operators, and the resulting Bayesian analysis across representative charge assignments.
Section~\ref{sec:Bayesian_SSM_FN} presents the Bayesian inference of the SSM incorporating the FN mechanism.
In Section~\ref{sec:cosmology}, we briefly discuss additional implications from cosmological observations, such as those related to dark matter and reheating.
Finally, Section~\ref{sec:conclusion} summarizes our findings and outlines directions for future work.

\section{Setup: Supersymmetric Standard Model}
\label{sec:setup}
In this work, we consider a supersymmetric spectrum beyond the TeV scale, in which the sfermion masses are much heavier than the gaugino masses, which are of order~TeV. The Higgsino is also assumed to be of the same order as the sfermion masses. Such a spectrum is well motivated both by theoretical considerations of supersymmetry breaking and by the absence of superpartners in current collider experiments.

\subsection{Yukawa Interaction}
\label{sec:Yukawa}
We extend the SSM by introducing three gauge-singlet right-handed neutrino superfields $\bar{N}_a$ ($a=1,2,3$) with $R$-parity conservation.  
The superpotential in this framework is given by  
\begin{align}
\label{eq:superpotential}
W =(y_u)_{ij} H_u Q_i \bar{u}_j
+(y_d)_{ij} H_d Q_i \bar{d}_j
+(y_e)_{ij} H_d  \bar{e}_i L_j
+(y_\nu)_{ia} H_u L_i \bar{N}_a
+\frac{1}{2} (M_N)_{ab} \bar{N}_a \bar{N}_b
+\mu H_u H_d\ ,
\end{align}
where $Q_i$ denotes the left-handed quark doublet chiral superfield,  
$\bar{u}_j$ and $\bar{d}_j$ the right-handed up- and down-type quark singlet superfields,  
$L_i$ the left-handed lepton doublet chiral superfield,  
$\bar{e}_j$ the right-handed charged lepton singlet superfield,  
and $H_{u,d}$ the Higgs doublet chiral superfields.  
The indices $i,j$ run over $1,2,3$.

The first three terms correspond to the standard Yukawa interactions of the SSM, with $y_u$, $y_d$, and $y_e$ denoting $3\times 3$ complex matrices in generation space. 
The fourth term introduces the neutrino Yukawa interactions, described by a $3\times 3$ complex matrix $y_\nu$, which couples the right-handed neutrinos $\bar{N}$ to the lepton doublets $L$ and the up-type Higgs field $H_u$.  
The fifth term represents the Majorana mass terms for the right-handed neutrinos, with $M_N$ denoting the heavy Majorana mass matrix:
\begin{align}
    (M_N)_{ab} &= M_{N,0} \, (c_N)_{ab} \ , \label{eq:MNc}
\end{align}
where $(c_N)_{ab}$ is a dimensionless complex symmetric matrix encoding the flavor structure, and $M_{N,0}$ is a dimensionful parameter setting the overall mass scale.
The last term is the so-called $\mu$-term, which couples the two Higgs doublets $H_u$ and $H_d$.

In the SSM, electroweak symmetry breaking is achieved by two Higgs doublets, $H_u$ and $H_d$, which acquire vacuum expectation values (VEVs) $v_u$ and $v_d$, respectively.  
These VEVs determine the relative sizes of the Yukawa couplings and thereby control the fermion masses,  
\begin{align}
(m_u)_{ij} = (y_u)_{ij}\, v_u \ , \quad (m_d)_{ij} = (y_d)_{ij}\, v_d \ , \quad (m_e)_{ij} = (y_e)_{ij}\, v_d \ .
\end{align}
Furthermore, 
the active neutrino masses are generated as  
\begin{align}
m_\nu = y_\nu v_u^2 M_N^{-1} y_\nu^T \ ,
\end{align}  
through the seesaw mechanism~\cite{Minkowski:1977sc,Yanagida:1979as,Yanagida:1979gs,Gell-Mann:1979vob,Glashow:1979nm,Mohapatra:1979ia}.  

It is customary to parametrize the ratio of the two Higgs VEVs as  
\begin{align}
\tan\beta = \frac{v_u}{v_d}\ ,
\end{align}
where $\tan\beta$ is one of the central parameters of the SSM.  
A crucial requirement for the viability of this framework is the successful reproduction of the observed Higgs boson mass.  
In PeV-scale supersymmetry, radiative corrections to the Higgs quartic coupling can naturally yield the measured value of $m_h \simeq 125~\mathrm{GeV}$~\cite{Okada:1990vk,Okada:1991jc,Ellis:1990nz,Haber:1990aw,Ellis:1991zd}.  
To this end, we adopt $\tan\beta = 2.5$ as a benchmark choice, which consistently reproduces the Higgs mass.  
This benchmark will serve as the reference point for our subsequent analysis of flavor structures and proton decay in supersymmetric models, while the sfermion mass scale is treated as a free parameter

The Yukawa matrices are generally non-diagonal in the general basis. 
To obtain the physical fermion masses, we perform singular value decomposition
to obtain the diagonal Yukawa matrices 
\begin{align}
\label{eq:Uu}
y_u &= U_{Q_u}^T\, y_u^{\mathrm{diag}}\, U_{\bar{u}}\ , \\
\label{eq:Ud}
y_d &= U_{Q_d}^T\, y_d^{\mathrm{diag}}\, U_{\bar{d}}\ , \\
\label{eq:Ue}
y_e &= U_{\bar{e}}^T\, y_e^{\mathrm{diag}}\, U_L\ ,
\end{align}
where $U$ and $V$ are unitary matrices acting on the left- and right-handed fermions, respectively. 
In the diagonal Yukawa basis, the CKM matrix $V_{CKM}$ is given by
\begin{align}
    V_\mathrm{CKM} = U_{Q_u} U_{Q_d}^\dagger \ ,
\end{align}
up to unphysical phase factors.

Similarly, in the neutrino sector, the light neutrino mass matrix obtained through the seesaw mechanism is a complex symmetric matrix. 
It can be diagonalized by the Takagi decomposition,  
\begin{align}
\label{eq:Unu}
m_\nu = U_\nu^T\, m_\nu^{\mathrm{diag}}\, U_\nu\ ,
\end{align}
where $U_\nu$ is a unitary matrix and $m_\nu^{\mathrm{diag}}$ contains the physical neutrino masses.  
In the charged-lepton diagonal basis, the PMNS matrix is then given by  
\begin{align}
    U_\mathrm{PMNS} = U_L U_\nu^\dagger\ ,
\end{align}
up to unphysical phase factors.

\subsection{Soft Masses and Higgsino Mass}

We now turn to the discussion of the soft supersymmetry-breaking mass terms.  
In our scenario, the characteristic soft mass scale is taken to be far above the electroweak scale.  
As a consequence, the effects of electroweak symmetry breaking on the scalar mass spectrum are negligible.  
We therefore disregard terms proportional to the Higgs VEVs and focus on the dominant contributions arising from the high-scale supersymmetry-breaking sector.

In our analysis, we consider mini-split supersymmetry in a generic setting,  
where the sfermion and gaugino masses are hierarchically separated~(see e.g., Refs.\,\cite{Hall:2011jd, Hall:2012zp, Nomura:2014asa, Ibe:2011aa, Ibe:2012hu, Arvanitaki:2012ps, ArkaniHamed:2012gw}).  
For the gaugino sector, we fix the masses as  
\begin{align}
M_1 = 6~\mathrm{TeV}\ , \quad M_2 = 3~\mathrm{TeV}\ , \quad M_3 = -20~\mathrm{TeV}\ .
\end{align}
This choice is motivated by the anomaly-mediated supersymmetry breaking (AMSB) mechanism~\cite{Randall:1998uk, Giudice:1998xp}, in which, in particular, the thermal relic abundance of the Wino can account for the observed dark matter density.  
We also assume that the soft trilinear supersymmetry-breaking parameters are of the same order as the gaugino masses, as suggested by the AMSB framework, and therefore neglect them in the following analysis.

In our framework, scalar soft masses are generated via higher-dimensional operators in the K\"{a}hler potential.  
A representative form of such an operator is  
\begin{align}
K \;\supset\; \frac{(c_{\phi})_{ij}}{M_*^{\,2}} \; X^\dagger X \; \phi_i^\dagger \phi_j \  ,
\end{align}
where $X$ denotes a hidden-sector chiral superfield whose $F$-component, $F_X$, acquires a nonzero VEV and thereby breaks supersymmetry.  Here, 
$\phi_i$ denotes a visible-sector matter chiral superfield, and $M_*$ is the characteristic mediation scale suppressing the operator, for example, the Planck scale in gravity mediation.  
The dimensionless coefficients $(c_{\phi})_{ij}$ encode the flavor dependence of the coupling between the hidden and visible sectors, and in general form a $3 \times 3$ Hermitian matrix.

When supersymmetry is broken by a nonzero $F$-term, $F_X \neq 0$, the above interaction generates soft scalar masses in the visible sector of the form
\begin{align}
\left(m_{\widetilde{f}}^2\right)_{ij} \;=\; (c_{\phi})_{ij} \, \frac{|F_X|^2}{M_*^{\,2}}  = m_0^2  (c_{\phi})_{ij}\ ,
\end{align}
where $m_{\widetilde{f}}^2$ denotes the sfermion mass-squared matrices and $m_0$ is their characteristic scale.
When supergravity effects are taken into account, there can also be contributions to the universal scalar mass.
In the present analysis, however, we neglect such contributions for simplicity.

We now move to the basis in which the fermion Yukawa matrices are diagonal.  
In this basis, the sfermion mass-squared matrices are transformed by the same unitary rotations that diagonalize the corresponding fermion fields 
\begin{align}
\widetilde{m}_{\widetilde{f}}^{\,2} &= U_{f}\, m_{\widetilde{f}}^{\,2} \, U_{f}^{\dagger} \ ,
\end{align}
where $U_f$ denotes the fermion diagonalization matrix for the species $f$ in Eqs.\,\eqref{eq:Uu},
\eqref{eq:Ud},
\eqref{eq:Ue}, and 
\eqref{eq:Unu}.  
At this stage, the matrices $\widetilde{m}_{\widetilde{f}}^{\,2}$ are generically non-diagonal,  
reflecting the misalignment between the fermion and sfermion sectors.

The next step is to diagonalize these $3\times 3$ Hermitian sfermion mass-squared matrices:
\begin{align}
\hat{m}_{\widetilde{f}}^{\,2}
&=
\operatorname{diag}\bigl(
m_{\widetilde{f}_1}^{\,2},\,
m_{\widetilde{f}_2}^{\,2},\,
m_{\widetilde{f}_3}^{\,2}
\bigr)
=
R_{\widetilde{f}}\,
\widetilde{m}_{\widetilde{f}}^{\,2}\,
R_{\widetilde{f}}^{\dagger} \  ,
\end{align}
where $R_{\widetilde{f}}$ denotes the unitary matrix that diagonalizes the sfermion mass matrix in flavor space.  
The eigenvalues $\hat{m}_{\widetilde{f}}^{\,2}$ correspond to the physical sfermion masses,  
while the mixing matrices $R_{\widetilde{f}}$ encode the residual flavor violation that appears in the couplings of gauginos and Higgsinos to matter fields.

In this work, we assume that the supersymmetric $\mu$-term is of the same order as the sfermion mass scale, namely
\begin{align}
\left| \mu \right| = \order{m_0} \  .
\end{align}
In general, $\mu$ is a complex quantity, and in our analysis we make no restriction on its complex phase; it is treated as a free parameter.  
The Higgs soft masses, $m_{H_u}^2$ and $m_{H_d}^2$, as well as the $b$-term, are chosen such that electroweak symmetry breaking is successfully achieved.
However, the specific choice of these parameters does not significantly affect the following discussion.

In the setup considered in this paper, collider constraints are not directly applicable.  The strongest current bound on a Wino LSP arises from searches for disappearing charged tracks produced by the charged Winos.  
The present LHC limit is approximately $660\,\mathrm{GeV}$ \cite{2201.02472,2309.16823}. 
Although the lifetime of the charged Wino can be affected by the Higgsino mass, for Higgsino masses above about $10\,\mathrm{TeV}$ the impact on this bound is negligible \cite{2210.16035,2312.08087}.

\subsection{Dimension-Five Proton Decay Operators}
Throughout this paper, we assume $R$-parity conservation, which forbids the renormalizable operators that violate baryon~number.  
However, even with $R$-parity conservation, the following non-renormalizable dimension-five operators responsible for proton decay are not forbidden.
They can be written as
\begin{align}
\mathcal{L}_{\mathrm{eff}}^{(5)} 
= \frac{C^{5L}_{ijkl}}{\Lambda_B}\, \mathcal{O}^{5L}_{ijkl} 
+ \frac{C^{5R}_{ijkl}}{\Lambda_B}\, \mathcal{O}^{5R}_{ijkl} 
+ \text{h.c.}\ ,
\end{align}
where
\begin{align}
\mathcal{O}^{5L}_{ijkl} &\equiv 
\int d^2\theta \;
\frac{1}{2}\, \epsilon^{abc}\,
\big(Q^a_i \cdot Q^b_j\big)\,
\big(Q^c_k \cdot L_l\big)\, , \\
\mathcal{O}^{5R}_{ijkl} &\equiv 
\int d^2\theta \;
\epsilon^{abc}\,
\bar{u}^a_i \bar{e}_j\,
\bar{u}^b_k \bar{d}^c_l\ .
\end{align}
Here $a,b,c$ denote color indices and $i,j,k,l$ label the flavor indices. 
The dimensionless coefficients $C^{5L}_{ijkl}$ and 
$C^{5R}_{ijkl}$ are complex valued, 
and $\Lambda_B$ denotes the dimensionful parameter.
The operators $\mathcal{O}^{5L}$ and $\mathcal{O}^{5R}$ correspond to dimension-five baryon-number violating terms, respectively. They arise, for example, from the exchange of color-triplet Higgs superfields in minimal supersymmetric GUTs such as $\mathrm{SU}(5)$, after integrating out the heavy triplet states.

To connect with the super-CKM basis, 
where up quark and lepton superfields are rotated to diagonalize the Yukawa matrices. 
In this basis, the Wilson coefficients 
are given by,
\begin{align}
\hat{C}^{5L}_{ijkl} &= 
(U_{Q_u}^\ast)_{ii'}\,(U_{Q_u}^\ast)_{jj'}\,
(U_{Q_u}^\ast)_{kk'}\,(U_{L}^\ast)_{ll'}\,
C^{5L}_{i'j'k'l'} \, , \\
\hat{C}^{5R}_{ijkl} &= 
(U_{\bar{u}}^\ast)_{ii'}\,(U_{\bar{e}}^\ast)_{jj'}\,
(U_{\bar{u}}^\ast)_{kk'}\,(U_{\bar{d}}^\ast)_{ll'}\,
C^{5R}_{i'j'k'l'} \ ,
\end{align}
where the unitary matrices are defined in 
Subsec.~\ref{sec:Yukawa}.

\section{Proton Decay and Flavor/CP Observation}
\label{sec:proton_flavor}

\subsection{Proton Decay}

\begin{figure}[t]
  \centering
   \begin{tikzpicture}[line cap=round,line join=round]
  \tikzset{
    thickline/.style={line width=1.1pt},
    dline/.style={dash pattern=on 3pt off 3pt, line width=1.1pt}
  }

  \coordinate (TopV)  at (0,2.2);
  \coordinate (Ltop)  at (-2.1,3.0);
  \coordinate (Rtop)  at ( 2.1,3.0);

  \coordinate (Ltri)  at (-1.1,0.6);
  \coordinate (Rtri)  at ( 1.1,0.6);

  \coordinate (Lbot)  at (-1.7,0.0);
  \coordinate (Rbot)  at ( 1.7,0.0);
  \coordinate (Lext)  at (-2.4,-1.0);
  \coordinate (Rext)  at ( 2.4,-1.0);
  \coordinate (Mid)   at (0,0);

  \draw[thickline] (Ltop) -- (TopV);
  \draw[thickline] (Rtop) -- (TopV);

  \draw[dline] (TopV) -- (Lbot);
  \draw[dline] (TopV) -- (Rbot);

  \draw[thickline] (Lbot) -- (Rbot);

  \draw[thickline] (Lext) -- (Lbot);
  \draw[thickline] (Rext) -- (Rbot);

  \filldraw[fill=gray!30, draw=black] (TopV) circle (0.16); 
  \fill[black] (Mid) circle (0.12);          
  \node at (0,-0.55) {$\widetilde g,\,\widetilde W,\,\widetilde B,\,\widetilde H_{u,d}$};

  \node at (-1.2,1.5) {$\widetilde \ell, \, \widetilde q$};
  \node at (0,2.7) {$\mathcal{O}^{5L,R}$};

\end{tikzpicture}
  \caption{Effective dimension-six baryon-number violating operators generated by dressing the dimension-five baryon-number violating operators by the sfermion-gaugino/Higgsino loops.
  The gray blob denotes the dimension-five operator and black dot represents the mass insertion of the gauginos or Higgsinos.
 }
   \label{fig:dressing}
\end{figure}
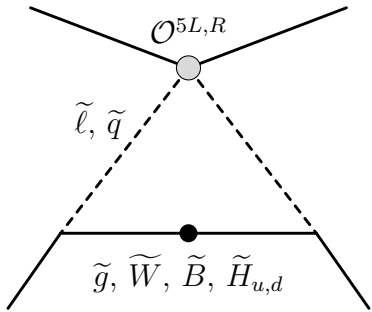

In this paper, we do not assume a specific GUT framework.  
Nevertheless, in general one expects the existence of the dimension-five baryon-number violating effective operators introduced above.  
Such superpotential operators can be ``dressed'' by sfermion-gaugino/Higgsino loops, resulting in effective dimension-six four-fermion operators that mediate proton decay~\cite{Weinberg:1979sa}  (see Fig.\,\ref{fig:dressing}).  
The mass insertion (chirality flip) along the internal higgsino/gaugino lines appears explicitly in the decay amplitude,
\begin{align}
\text{Higgsino dressing:} \quad 
 &\simprop 
y^2 \, \frac{\mu}{m_{\widetilde{f}}^{\,2}} \, \frac{1}{\Lambda_B} \ , \\[4pt]
\text{Gaugino dressing:} \quad 
 &\simprop 
g^2 \, \frac{M_i}{m_{\widetilde{f}}^{\,2}} \, \frac{1}{\Lambda_B} \ .
\end{align}

From these parametric dependencies, it is clear that unless the Higgsino mass $\mu$ is sufficiently larger than the gaugino masses $M_i$, the gaugino contribution dominates the dressing.  
Among the gauginos, the gluino contribution is typically enhanced due to its large gauge coupling and, in our setup, its large mass parameter.

In models where the sfermion mass structure respects minimal flavor violation (MFV),  only amplitudes picking up CKM mixing from Higgsino or Wino exchange survive, and the gluino contribution is suppressed~\cite{Belyaev:1982ik}.  
In contrast, in the present study we do not impose MFV, but rather consider a generic flavor structure for the sfermion sector.  
In this situation, the gluino-mediated dressing diagrams can yield sizable contributions to the proton decay amplitudes, often dominating over electroweak gaugino and Higgsino channels.

For the calculation of proton decay rates, we employ the standard formalism in which the partial decay width is obtained from the Wilson coefficients of the effective dimension-five operators, evolved to the hadronic scale and combined with the corresponding hadronic matrix elements.  
For the nucleon matrix elements, we use the central values obtained from recent lattice QCD calculations in Refs.\,\cite{Yoo:2021gql,Aoki:2017puj}.  
The resulting decay rates are compared with the current experimental limits from Super-Kamiokande\,\cite{Super-Kamiokande:2020wjk,Super-Kamiokande:2013rwg,Super-Kamiokande:2005lev,Super-Kamiokande:2022egr,Super-Kamiokande:2014otb,Super-Kamiokande:2024qbv}, and with the projected sensitivities of Hyper-Kamiokande\,\cite{Hyper-Kamiokande:2018ofw}.

\subsection{Flavor/CP Observables}
In general, the sfermion mass matrices can violate flavor symmetries and CP invariance.  
Such violations give rise to FCNCs and CP-violating observables that deviate from the SM predictions.  
Consequently, precision measurements of CP violation and FCNC processes at low energies provide powerful indirect probes of the sfermion sector.  

\begin{figure}[t]
    \centering
    \subcaptionbox{$\varDelta F=2$ processes\label{fig:box}}[0.45\textwidth]{\begin{tikzpicture}
  \coordinate (Ex) at (2.4, 0);
  \coordinate (Ey) at (0, 1.6);
  \coordinate (Bx) at (1.0, 0);
  \coordinate (By) at (0, 1.0);
  \begin{feynhand}
    \vertex (L1) at ($-1*(Ex)+(Ey)$) {$\bar q_i$};
    \vertex (L2) at ($-1*(Ex)-(Ey)$) {$q_i$};
    \vertex (R1) at ($(Ex)+(Ey)$) {$\bar q_j$};
    \vertex (R2) at ($(Ex)-(Ey)$) {$q_j$};

    \vertex (A) at ($-1*(Bx)+(By)$);
    \vertex (B) at ($(Bx)+(By)$);
    \vertex (C) at ($(Bx)-(By)$);
    \vertex (D) at ($-1*(Bx)-(By)$);

    \graph{
      (L1) -- [plain,thick] (A),
      (D)  -- [plain,thick] (L2),

      (B)  -- [plain,thick] (R1),
      (R2) -- [plain,thick] (C),

      (B) -- [plain,thick, edge label'={\(\widetilde g\)}] (A),
      (D) -- [plain,thick, edge label'={\(\widetilde g\)}] (C),

      (A) -- [sca,thick, edge label'={\(\widetilde q\)}] (D),
      (B) -- [sca,thick, edge label={\(\widetilde q\)}] (C)
    };
  \end{feynhand}
\end{tikzpicture}}
    \subcaptionbox{EDM\label{fig:EDM}}[0.4\textwidth]{\begin{tikzpicture}[scale=1.,
  fermion/.style={thick, postaction={decorate},
    decoration={markings,
      mark=at position 0.25 with {\arrow{>}},
      mark=at position 0.75 with {\arrow{<}}}},
   exfermion/.style={thick, postaction={decorate},
    decoration={markings,
      mark=at position 0.5 with {\arrow{>}}}}, 
   scalar/.style={thick, dashed},
  photon/.style={decorate, decoration={snake, amplitude=1.5pt, segment length=6pt}, thick},
  dot/.style={circle, fill=black, inner sep=1.1pt},
]
  \def\R{1.25}        
  \coordinate (L) at (-3,0);
  \coordinate (R) at ( 3,0);
  \coordinate (A) at (-\R,0);
  \coordinate (B) at ( \R,0);
  \coordinate (O) at (0,0);

  \draw[exfermion] (A)--(L);
  \draw[exfermion] (B)--(R);

  \node[dot] at (A) {};
  \node[dot] at (B) {};

  \draw[fermion] (A) arc[start angle=180,end angle=0,radius=\R cm];
  \node[] at (0,1.3*\R) {$\widetilde{g}, \widetilde{B}$};

\filldraw[fill=black, draw=black]  (0,\R) circle (0.12);

  \draw[scalar] (A) arc[start angle=180,end angle=360,radius=\R cm];

  \foreach \ang in {-150,-30}{
    \coordinate (P) at ($(O)+({\R*cos(\ang)},{\R*sin(\ang)})$);
    \draw[line width=0.6pt] ($(P)+(-0.12,0)$) -- ++(0.24,0);
    \draw[line width=0.6pt] ($(P)+(0,-0.12)$) -- ++(0,0.24);
  }
    \draw[line width=0.6pt] ( -0.09,-\R -0.09) -- ++(0.18,0.18);
    \draw[line width=0.6pt] ( -0.09,-\R +0.09) -- ++(0.18,-0.18);
  \coordinate (E) at ($(O)+({1.2*\R*cos(-50)},{-1.2*\R*sin(-50)})$);
  \draw[photon] (E) -- ++(1.0,0.8);
  \node[below right] at ($(E)+(1.0,0.8)$) {$\gamma,\, g$};

  \node at (-1.2*\R,-0.4*\R) {$\widetilde{u}_R$};
  \node at ( 1.2*\R,-0.4*\R) {$\widetilde{u}_L$};
  \node at (-0.5*\R,-1.1*\R) {$\widetilde{t}_R$};
  \node at (0.5*\R,-1.1*\R) {$\widetilde{t}_L$};
  
\end{tikzpicture}}
    \caption{Flavor- and CP-violating processes arising from squark mass matrices.}
\end{figure}
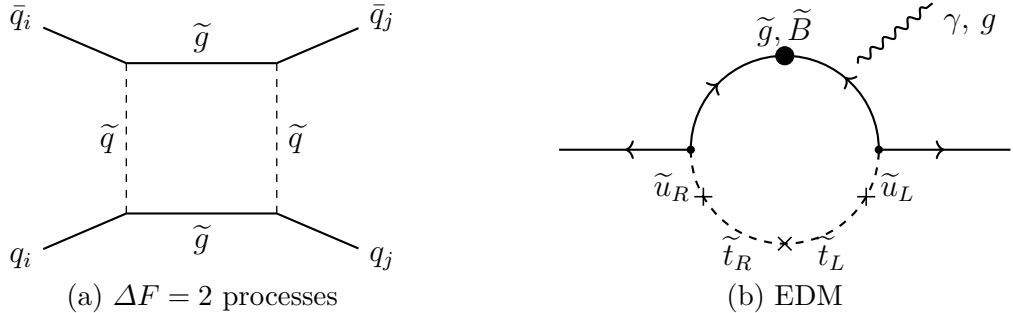

Among the most stringent constraints come from $\varDelta F = 2$ transitions,  
such as the precise measurements of neutral meson mixing, as well as from the searches for fermion electric dipole moments (EDMs).  
These observables are highly sensitive to the off-diagonal and complex components of the sfermion mass matrices,  
and therefore impose tight limits on the possible flavor and CP structure of supersymmetric models.
In the present analysis, we consider the following neutral meson mixing systems as probes of flavor violation:
\begin{align}
 K^0\!-\!\bar{K}^0\ , \quad D^0\!-\!\bar{D}^0\ , \quad B_d^0\!-\!\bar{B}_d^0\ , \quad B_s^0\!-\!\bar{B}_s^0 \ .   
\end{align}
These processes constrain $\varDelta F = 2$ transitions and can receive new-physics contributions from box diagrams involving sfermion-gaugino exchange (see Fig.~\ref{fig:box}).  
From such box diagrams, we compute the effective quark four-fermion operators and match them onto the low-energy meson mixing Hamiltonian.

For the hadronic matrix elements, we employ lattice QCD results as follows:  
for the $K^0$ system we use the values from Ref.\,\cite{2404.02297},  
for the $B_d^0$ and $B_s^0$ systems from Ref.\,\cite{1907.01025},  
and for the $D^0$ system from Ref.\,\cite{1706.04622}.  
The experimental bounds are taken from the UTfit 2023 new-physics analysis for $K$ and $B$ mesons \cite{UTfit:2022hsi}, and from the HFLAV results given in Ref.\,\cite{2411.18639} for $D$ mesons.

In addition to meson mixing, we consider the EDMs of the neutron and the electron as sensitive probes of CP violation.  
Since these EDMs arise from chirality-flipping processes, their amplitudes receive contributions proportional to the Higgsino mass $\mu$ or the gaugino mass parameters $M_i$ (see Fig.~\ref{fig:EDM}).  
Moreover, sfermion flavor violation can enhance chirality-flip effects via loops involving heavy fermions, thereby increasing the experimental sensitivity.
In the present analysis, we consider only the one-loop diagrams involving sfermions.

For the electron and neutron EDMs, we adopt the current experimental limits reported in Refs.\,\cite{2212.11841, Abel:2020pzs}.  
In the evaluation of the neutron EDM, both quark EDM and quark chromo-EDM contributions are included, and the total value is computed following the methodology of Ref.\,\cite{2303.02822}.

\subsection{Bayesian Approach}

In this subsection, we examine how the constraints from proton decay, as well as from the FCNC and CP-violating observables discussed above, restrict the sfermion sector of the supersymmetric standard model.  
In the absence of underlying symmetry or dynamical mechanism imposing additional structure, it is natural to expect that the dimensionless Lagrangian parameters,  such as $y$ in Yukawa couplings, $c_{\phi}$ in the sfermion mass matrices and $C^{5L,R}$ in the dimension-five baryon-number violating operators, are generic complex numbers of $\mathcal{O}(1)$.

We adopt this assumption as the basis for our prior distributions in the Bayesian analysis.  
Specifically, we take the $3\times 3$ Hermitian matrix $c_{\phi}$ to be distributed as
\begin{equation}
(c_{\phi})_{ij} \;\sim\;
\begin{cases}
\mathcal{N}(0,1) & \quad (i=j)\ , \\[4pt]
\mathcal{N}\!\left(0,\tfrac{1}{\sqrt{2}}\right)
+ i\,\mathcal{N}\!\left(0,\tfrac{1}{\sqrt{2}}\right) & \quad (i \neq j)\ .
\end{cases}
\end{equation}
Since $c_\phi$ determines the sfermion mass-squared matrices, we restrict our analysis to the positive-definite cases in which all eigenvalues of $m_{\widetilde{f}}^2$ are strictly positive.  
This guarantees the absence of tachyonic sfermion states and ensures a physically viable scalar spectrum.
The complex coefficients $C^{5L}, C^{5R}$ and $y$ are taken as
\begin{align}
y,\, C^{5L,R}  &\sim \mathcal{N}\!\left(0,\tfrac{1}{\sqrt{2}}\right)
+ i\,\mathcal{N}\!\left(0,\tfrac{1}{\sqrt{2}}\right) \ ,
\end{align}
and complex symmetric matrix,
\begin{align}
    (c_{N})_{ab}  \;\sim\;
\begin{cases}
\mathcal{N}\!\left(0,\tfrac{1}{\sqrt{2}}\right) + i\mathcal{N}\!\left(0,\tfrac{1}{\sqrt{2}}\right) & \quad (i=j)\ , \\[4pt]
\mathcal{N}\!\left(0,\tfrac{1}{2}\right)
+ i\,\mathcal{N}\!\left(0,\tfrac{1}{2}\right) & \quad (i \neq j)\ .
\end{cases}
\end{align}
Here, $\mathcal{N}(\mu,\sigma)$ denotes a Gaussian distribution with mean $\mu$ and standard deviation $\sigma$.  
With these choices, the variance of the modulus of each complex parameter is unity.  
The motivation for adopting such $\mathcal{O}(1)$ priors is discussed in Ref.\,\cite{2412.19484}.

Using these priors, we compute the marginalized likelihood associated with proton decay and FCNC/CP constraints and all Yukawa matrices to quantify the statistical preference for different regions in the sfermion-sector parameter space. Specifically, we evaluate the likelihood by constructing a $\chi^2$ function from the experimental central values and the corresponding $1\sigma$ theoretical and experimental uncertainties for each observable. The marginalized likelihood, or Bayesian evidence, is defined as
\begin{align}
\label{eq: marginalized likelihood}
B(m_{0}, \Lambda_{B}; L_{\mathrm{obs}})
&= \int  d\boldsymbol{c}
\pi(\boldsymbol{c})\, L_{\mathrm{obs}}(\boldsymbol{c}) \ ,
\end{align}
where $\boldsymbol{c} = \{c_{\phi},\, C^{5L},\, C^{5R},\, y,\, c_{N}\}$ collectively denotes the set of complex $\mathcal{O}(1)$ coefficients, and $y$ includes all Yukawa matrices $y_{u}$, $y_{d}$, $y_{e}$, and $y_{\nu}$. The prior distribution $\pi(\boldsymbol{c})$ and the likelihood function $L_{\mathrm{obs}}(\boldsymbol{c})$ are defined as in the previous subsection, with the latter depending on observables relevant to each analysis, such as fermion Yukawa couplings, flavor-changing and CP-violating observables, and proton decay constraints.
The resulting marginalized likelihood quantifies how well each parameter set reproduces the observed data across these sectors.
The integration is performed via Monte Carlo methods, following the numerical procedure described in Ref.\,\cite{2412.19484}.

For the quark and charged-lepton sectors, we adopt the $\overline{\mathrm{DR}}$ Yukawa couplings and CKM parameters evaluated at $10^{16}\,\mathrm{GeV}$, obtained by renormalization group evolution in a supersymmetric spectrum with $m_{0} = \mu = 1~\mathrm{PeV}$ and $\tan\beta = 2.5$.  
The input values are based on the PDG averages of the SM particle masses~\cite{ParticleDataGroup:2022pth}, with the running parameters determined following Ref.\,\cite{Buttazzo:2013uya}.  
For the light quarks, QCD four-loop RGEs and three-loop decoupling corrections from heavy quarks are included~\cite{Chetyrkin:1997sg}.  
The central values and $1\sigma$ uncertainties used in the likelihood are
\begin{gather*}
y_{u} = (3.12 \pm 0.39)\times 10^{-6}\ , \quad
y_{c} = (1.58 \pm 0.04)\times 10^{-3}\ , \quad
y_{t} = 0.526 \pm 0.004\ ,\\
y_{d} = (1.51 \pm 0.10)\times 10^{-5}\ , \quad
y_{s} = (3.03 \pm 0.18)\times 10^{-4}\ , \quad
y_{b} = (1.60 \pm 0.02)\times 10^{-2}\ ,\\
y_{e} = (6.180 \pm 0.012)\times 10^{-6}\ , \quad
y_{\mu} = (1.305 \pm 0.003)\times 10^{-3}\ , \quad
y_{\tau} = (2.210 \pm 0.006)\times 10^{-2}\ ,\\
s_{12}^{\mathrm{CKM}} = 0.2250 \pm 0.0007\ , \quad
s_{23}^{\mathrm{CKM}} = 0.0421 \pm 0.0008\ , \quad
s_{13}^{\mathrm{CKM}} = 0.00371 \pm 0.00011\ ,\\
\delta_{\mathrm{CP}}^{\mathrm{CKM}} = 1.144 \pm 0.026 \ ,
\end{gather*}
evaluated at the GUT scale.
These parameters are used to construct the Yukawa-sector likelihood $L_{\mathrm{Yukawa}}$ in the Bayesian analysis.

For the neutrino sector, we employ observables that are insensitive to the absolute mass scale of right-handed neutrinos.  
Specifically, we use the ratio of mass-squared differences $\varDelta m_{12}^2 / \varDelta m_{13}^2$, the PMNS mixing angles $(s_{12}^{\mathrm{PMNS}})^2$, $(s_{23}^{\mathrm{PMNS}})^2$, $(s_{13}^{\mathrm{PMNS}})^2$, and the Dirac CP phase $\delta_{13}^{\mathrm{PMNS}}$.  
The adopted central values and $1\sigma$ uncertainties are
\begin{gather*}
\frac{\varDelta m_{12}^2}{\varDelta m_{13}^2} = (2.96 \pm 0.09)\times 10^{-2}\ , \quad
(s_{12}^{\mathrm{PMNS}})^2 = 0.304 \pm 0.012\ , \\
(s_{23}^{\mathrm{PMNS}})^2 = 0.450 \pm 0.018\ , \quad
(s_{13}^{\mathrm{PMNS}})^2 = 0.0225 \pm 0.0001\ , \quad
\delta_{13}^{\mathrm{PMNS}} = 4.01 \pm 0.52\ .
\end{gather*}
These values correspond to the fit results for the normal mass ordering reported by NuFit~5.2~\cite{Esteban:2020cvm}, and are used to constrain the lepton-mixing component of the Yukawa-sector likelihood $L_{\mathrm{Yukawa}}$.

It should be noted that the $\overline{\mathrm{DR}}$ Yukawa couplings at the high scale generally depend on details of the superpartner spectrum, such as the sfermion masses, flavor structure, and $\tan\beta$.  
However, for simplicity and to maintain a consistent reference across all benchmark scenarios, we fix the $\overline{\mathrm{DR}}$ values to those listed above throughout this analysis.

To study the constraints on the sfermion mass scale, we consider the normalized Bayes factor
\begin{align}
\mathrm{BF}(m_0, \Lambda_B ; L_{\mathrm{obs}}) \;=\; 
\frac{B(m_0, \Lambda_B ; L_{\mathrm{obs}})}{B(m_0 \to \infty, \Lambda_B ; L_{\mathrm{obs}})} \ .
\end{align}
In the limit $m_0 \to \infty$, the sfermions effectively decouple from low-energy processes.  
In this regime, any choice of sfermion or dimension-five operator structure is consistent with current experimental bounds, 
and hence, $B(m_0 \to \infty, \Lambda_B ; L_{\mathrm{obs}})$ serves as a reference likelihood corresponding to completely unconstrained high-scale supersymmetry scenario.
If one performs a Bayesian parameter estimation over $(m_0,\Lambda_B)$, the resulting posterior distribution is proportional to the product of the prior and this Bayes factor, $\pi(m_0,\Lambda_B)\, \mathrm{BF}(m_0,\Lambda_B)$.

Crudely speaking, the ratio $\mathrm{BF}(m_0, \Lambda_B)$ quantifies the fraction of configurations 
in the parameter space of $(c_{\phi}, C^{5L}, C^{5R})$ that are compatible with the observed  FCNC, CP-violating, and proton decay constraints, for a given finite sfermion mass scale $m_0$ (see Refs.\,\cite{Jeffreys:1939xee,Kass:1995loi}).  
In other words, it measures how much of the generic $\mathcal{O}(1)$ coefficient space remains consistent with experimental data once sfermions are brought down to a particular scale 
compared to the baseline case where they are arbitrarily heavy.

\begin{figure}[t]
\centering
\begin{subfigure}{0.47\textwidth}
  \centering
  \includegraphics[width=\linewidth]{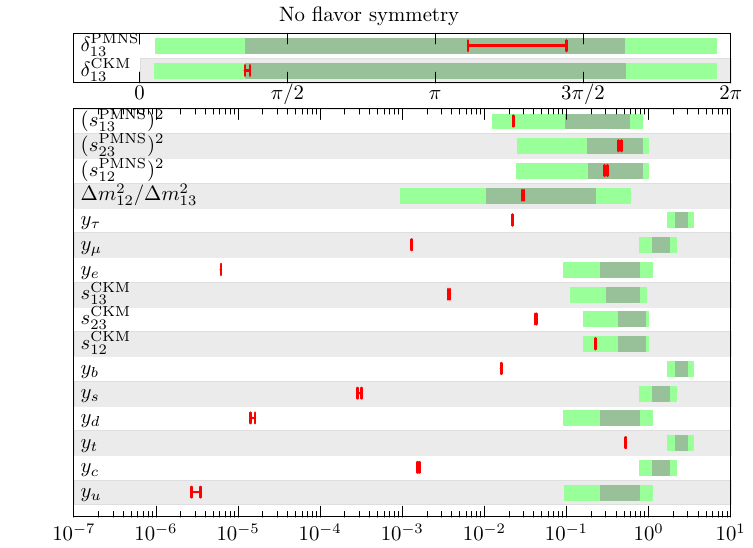}
\caption{Prior distributions of the Yukawa couplings and fermion mixings}  
  \label{fig:NcaseYukawa}
\end{subfigure}
\begin{subfigure}{0.47\textwidth}
  \centering
  \includegraphics[width=\linewidth]{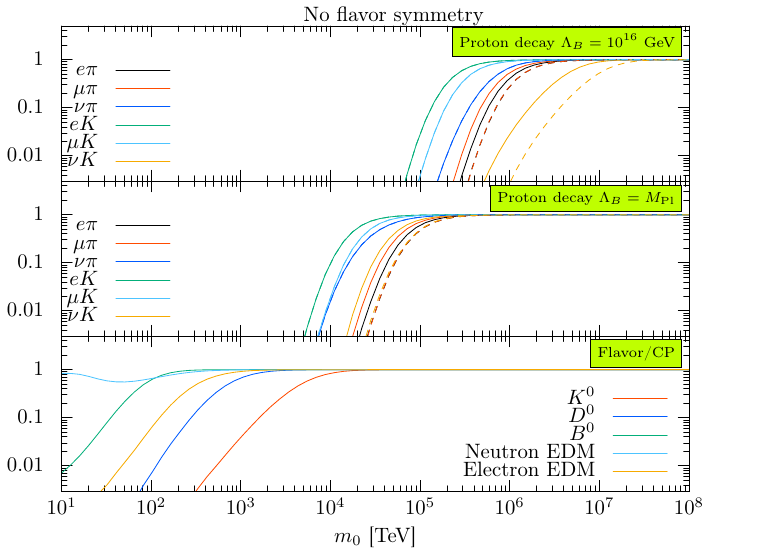}
   \caption{Bayes factor from flavor/CP and proton decay.} \label{fig:NcaseBayes}
\end{subfigure}
\caption{
(a) Prior distributions of the Yukawa couplings and fermion mixings.  The shaded regions correspond to the $1\sigma$ (light green) and $2\sigma$ (dark green) ranges predicted by the priors, with the measured values indicated for comparison.  
(b) Bayes factor as a function of $m_0$ for different experimental constraints, including FCNC bounds from meson mixing, fermion EDM limits, and proton-decay constraints for $\Lambda_B = 10^{16}\,\mathrm{GeV}$ and $\Lambda_B = M_{\mathrm{Pl}}$.  
Dashed lines represent the projected sensitivities of Hyper-Kamiokande.
}
\label{fig:Ncase}
\end{figure}

Figure~\ref{fig:NcaseYukawa} presents the prior distributions of the observables implied by the model-parameter priors discussed above, together with the corresponding Bayes factors. 
The figure shows the $1\sigma$ and $2\sigma$ ranges of the prior distributions obtained without including any likelihood information from the observational data, as well as the actual measured values. 
From this comparison, it is clear that the distribution predicted by  naive $\mathcal{O}(1)$ Yukawa couplings is strongly inconsistent with the experimentally observed parameters, that is, a mismatch commonly referred to as the \emph{fermion mass hierarchy} or \emph{flavor puzzle}.  
The Bayesian analysis for models that reproduce the observed SM flavor structure will be discussed in the next section.

Figure~\ref{fig:NcaseBayes} shows the Bayes factor as a function of the sfermion mass scale $m_{0}$.  
The likelihood used in this analysis combines  contributions from the Yukawa sector and the low-energy observables, 
\begin{align}
L_{\mathrm{obs}} = L_{\mathrm{Yukawa}} \times L_{\mathrm{FCNC/CPV/PD}} \ ,
\end{align}
where $L_{\mathrm{Yukawa}}$ constrains the fermion masses and mixings, and 
$L_{\mathrm{FCNC/CPV/PD}}$ denotes the likelihood associated with flavor-changing, CP-violating, or proton decay observables, depending on the analysis.  
For the flavor and CP observables, we employ the constraints from meson mixing and fermion EDMs discussed above.  
The upper two panels display the Bayes factors derived from proton decay constraints for two representative values of the baryon~number-violating scale: 
$\Lambda_{B} = 10^{16}\,\mathrm{GeV}$ and $\Lambda_{B} = M_{\mathrm{Pl}}$, where $M_{\mathrm{Pl}}$ denotes the reduced Planck mass ($2.4\times10^{18}\,\mathrm{GeV}$).  
In this analysis, we set $\abs{\mu} = m_{0}$.
Each line corresponds to the decay mode indicated in the figure.

The FCNC bounds are already extremely stringent: the constraint from $K$-meson mixing alone drives the Bayes factor below $0.1$ even for sfermion masses around $1\,\mathrm{PeV}$, effectively excluding large portions of the parameter space. 
Proton-decay limits are even more restrictive. 
Even with a Planck-scale cutoff, the experimental tension remains, indicating that sfermions as heavy as $10\,\mathrm{PeV}$ remain disfavored.

\section{Flavor Symmetry in SSM: Froggatt-Nielsen Approach}
\label{sec:FN}

As seen in the previous section, assuming dimensionless parameters to be simply $\mathcal{O}(1)$ fails to reproduce the observed Yukawa couplings. 
In the case of supersymmetry, an even more severe problem arises: proton decay imposes extremely strong constraints. 
These couplings must therefore be far more suppressed than $\mathcal{O}(1)$.

A common approach to achieve such suppression is to introduce flavor symmetries, which can control and reduce these unwanted $\mathcal{O}(1)$ parameters. 
In this paper, we focus on one of the simplest possibilities: the FN mechanism based on a U$(1)$ symmetry, U$(1)_{\mathrm{FN}}$ \cite{Froggatt:1978nt}.

\subsection{Fermion Mass and Mixing}
In the FN mechanism, the SSM chiral fields are assigned distinct U$(1)_{\mathrm{FN}}$ charges.  
The symmetry is spontaneously broken by the VEV of a flavon field $\Phi$ carrying charge $-1$.  
Because we work in a supersymmetric setup, we also introduce a chiral flavon field $\bar{\Phi}$ with charge $+1$, so that both $\Phi$ and $\bar{\Phi}$ can appear in the superpotential.  
For simplicity, we assume their VEVs to be equal,
\begin{equation}
\langle \Phi \rangle = \langle \bar{\Phi} \rangle\ .
\end{equation}

The U$(1)_{\mathrm{FN}}$ symmetry forbids unwanted $\order{1}$ Yukawa couplings 
at the renormalizable level.  
Once the symmetry is spontaneously broken, such interactions can arise through higher-dimensional operators suppressed by powers of the FN cutoff scale $\Lambda_{\mathrm{FN}}$.  
For fields with U$(1)_{\mathrm{FN}}$ charges $f_{H}$, $f_{\psi_i}$, and $f_{\chi_j}$, we define
\begin{equation}
q_{ij} \equiv \bigl|f_{H} + f_{\psi_i} + f_{\chi_j}\bigr|\ ,
\end{equation}
so that the leading operators appearing in the superpotential take the form
\begin{equation}
\begin{cases}
\displaystyle 
\left(\frac{\Phi}{\Lambda_{\mathrm{FN}}}\right)^{q_{ij}} H\,\psi_i\,\chi_j\ , 
& f_{H} + f_{\psi_i} + f_{\chi_j} > 0 \ , \\[1.0em]
\displaystyle 
\left(\frac{\bar{\Phi}}{\Lambda_{\mathrm{FN}}}\right)^{q_{ij}} H\,\psi_i\,\chi_j\ , 
& f_{H} + f_{\psi_i} + f_{\chi_j} < 0 \ .
\end{cases}
\end{equation}

This mechanism naturally generates an effective Yukawa structure.  
Introducing the FN breaking parameter
\begin{equation}
\epsilon \equiv 
\frac{\langle \Phi \rangle}{\Lambda_{\mathrm{FN}}} 
= \frac{\langle \bar{\Phi} \rangle}{\Lambda_{\mathrm{FN}}}\ ,
\end{equation}
the effective Yukawa couplings can be written schematically as
\begin{equation}
(y)_{ij} = \kappa^{(y)}_{ij}\,\epsilon^{\,q_{ij}}\ , 
\end{equation}
where $\kappa^{(y)}_{ij}$ are complex coefficients of order unity.  
For $\epsilon \sim 0.2$, comparable to the Cabibbo angle, hierarchical Yukawa structures naturally emerge from simple integer charge assignments.

This discussion can also be applied to the Yukawa couplings and masses of right–handed neutrinos.  
For the mass term, the dimensionless parameter $(c_{N})_{ab}$ appearing in Eq.\,\eqref{eq:MNc} is given by
\begin{equation}
(c_{N})_{ab} = \kappa^{(c_N)}_{ab}\,\epsilon^{\lvert f_{N_a} + f_{N_b} \rvert}\ ,
\end{equation}
where $\kappa^{(c_N)}_{ab}$ are expected to be complex parameters of order unity.  
The charges in the right–handed sector generally affect the seesaw mechanism as well.
Note that, in some cases the FN charges of the right-handed neutrinos do not affect the neutrino sector (ratio of mass–squared differences and mixing angles).  
For instance, if all chiral fields carry non–negative charges, the contributions from the right–handed charges cancel in the generation of light–neutrino masses via the seesaw mechanism, leaving the result independent of the right–handed neutrino charges.

\subsection{Sfermion Mass}

In supersymmetric frameworks, the FN mechanism not only generates hierarchical Yukawa couplings but also imprints a nontrivial structure on the soft supersymmetry-breaking scalar mass–squared matrices.  
Because the scalar fields carry $\mathrm{U}(1)_{\mathrm{FN}}$ charges, higher–dimensional operators in the K\"{a}hler potential with insertions of the flavon fields 
$\Phi/\Phi^{\dagger}$ and $\bar{\Phi}/\bar{\Phi}^{\dagger}$ induce off–diagonal terms suppressed by powers of the FN–breaking parameter $\epsilon$.  
Consequently, in the flavor basis the scalar mass–squared matrices take the schematic form
\begin{equation}
(m^2_{\widetilde f})_{ij} = m_0^2 
\kappa^{(c_\phi)}_{ij} \epsilon^{\lvert f_i - f_j\rvert} \ ,
\end{equation}
where $m_{0}$ is the typical scalar mass scale, $f_{i}$ denote the U$(1)_{\mathrm{FN}}$ charges of the corresponding superfields, and $\kappa^{(c_\phi)}_{ij}$ are ${\cal O}(1)$ coefficients.  
The diagonal entries remain unsuppressed, whereas the off–diagonal entries are naturally hierarchical, controlled by the charge differences among generations.  
Strictly speaking, whether the FN charges indeed govern the operators generating the sfermion masses is model dependent, and here we simply assume such a structure in the following discussion.

\subsection{Baryon Number Violating Dimension-Five Operators}

We now turn to the dimension-five operators responsible for proton decay.  
In this framework, the Wilson coefficients inherit the FN suppression factors through the Yukawa structure.  
Their schematic scaling is given by
\begin{align}
C^{5L}_{ijkl} &= \kappa^{(C^{5L})}_{ijkl}\,
\epsilon^{\lvert f_{Q_i}+f_{Q_j}+f_{Q_k}+f_{L_l} \rvert} \ , \\
C^{5R}_{ijkl} &= \kappa^{(C^{5R})}_{ijkl}
\epsilon^{\lvert f_{\bar{u}_i}+f_{\bar{e}_j}+f_{\bar{u}_k}+f_{\bar{d}_l} \rvert}\ ,
\end{align}
where $\kappa^{(C^{5L})}_{ijkl}$ and $\kappa^{(C^{5R})}_{ijkl}$ are ${\cal O}(1)$ coefficients and 
$f_{\psi}$ denote the $\mathrm{U}(1)_{\mathrm{FN}}$ charges of the corresponding fields.  
Thus, the FN charges not only reproduce the observed hierarchies of fermion masses but also control the relative strengths of baryon~number-violating operators.

It should be noted, however, that it is not guaranteed that the FN charges also control operators suppressed by the Planck scale.  
From the perspective of quantum gravity, exact global symmetries are expected not to exist, and in such cases the FN mechanism may fail to suppress these operators.  
In the following, we therefore also examine scenarios in which the baryon~number-violating dimension-five operators are not subject to FN suppression.

\subsection{Bayesian Analysis}
We now outline the Bayesian analysis in the presence of FN charges.  
The procedure follows the same steps as in the previous subsection, but in the FN framework the fundamental parameters of the SSM are replaced by the $\mathcal{O}(1)$ coefficients $\kappa$ that multiply the FN-suppressed operators.  
For a given charge assignment and the FN breaking parameter $\{f, \epsilon\}$, the corresponding marginalized likelihood, or Bayesian evidence, is obtained by modifying Eq.\,\eqref{eq: marginalized likelihood} to,
\begin{align}
B\bigl(m_{0}, \Lambda_{B}, \{f, \epsilon\}; L_{\mathrm{obs}}\bigr)
&= \int \! d\boldsymbol{\kappa}\;
\pi(\boldsymbol{\kappa})\, L_{\mathrm{obs}}\!\bigl(\boldsymbol{c}(\boldsymbol{\kappa}, \{f, \epsilon\})\bigr) \ .
\label{eq:BFFN}
\end{align}
Here, $\boldsymbol{\kappa} = \{\kappa^{(c_{\phi})},\, \kappa^{(C^{5L})},\, \kappa^{(C^{5R})},\, \kappa^{(y)},\, \kappa^{(c_{N})}\}$ collectively denotes the set of $\mathcal{O}(1)$ coefficients in the FN framework.  
The prior distribution $\pi(\boldsymbol{\kappa})$ is the same Gaussian ensemble introduced in the previous subsection.  
The likelihood $L_{\mathrm{obs}}\!\bigl(\boldsymbol{c}(\boldsymbol{\kappa}, \{f, \epsilon\})\bigr)$ is evaluated using the effective couplings 
$\boldsymbol{c} = \{c_{\phi}, C^{5L}, C^{5R}, y, c_N\}$ 
that are generated from the FN structure.  
Its functional form is identical to that used in the previous analysis, depending on the same set of observables such as fermion masses and mixings, flavor- and CP-violating observables, and proton decay constraints.

\subsection{Benchmark}
In this section, we present the benchmark charge assignments employed in our analysis.  
For simplicity, the $\mathrm{U}(1)_{\mathrm{FN}}$ charge of the $H_u$ field is set to zero throughout this work.  
Table~\ref{tab:FNcharges} summarizes the benchmark models $N$, $G$, $A$, $A_{\slashed{B}}$, $A'$, $A_{H_d(3)}$, $B$, $B'$, and $C$.  
For all benchmark models, we adopt the supersymmetric particle spectrum fixed in the previous section.

The benchmark models are defined as follows:
\begin{description}
  \item[Model $N$:] Scenario without any FN mechanism.
  \item[Model $G$:] FN charge assignments motivated by  GUT structures.
  \item[Model $A$:] Representative benchmark providing  excellent fits to the observed Yukawa couplings.
  \item[Model $A_{\slashed{B}}$:] FN charges identical to those of $A$, but without FN suppression for the baryon-number–violating dimension-five operators.
  \item[Model $A'$:] Same Yukawa structure as $A$, but with the signs of the lepton-sector charges reversed.
  \item[Model $A_{H_d(3)}$:] Defined such that $H_d$ carries an FN charge of $+3$, while $\bar{d}$ and $L$ are shifted by $-3$ relative to model $A$. This preserves the (s)lepton and (s)quark masses and mixings of $A$, but modifies the structure of the dimension-five operators.
  \item[Models $B$ and $C$:] Additional benchmarks that also provide  excellent fits to the observed Yukawa couplings.
  \item[Model $B'$:] Differs from $B$ only by the overall sign of the lepton-sector charges.
\end{description}
The supersymmetric $\mu$-term could, in principle, be subject to FN suppression depending on its origin.
In the present analysis, however, we fix its magnitude to be of order $m_0$.

The quantities $\mathrm{BF}_{q}$ and $\mathrm{BF}_{l}$ denote the Bayes factors defined as
\begin{align}
\mathrm{BF}_{q/l}
&= 
\frac{B\bigl(\{f, \epsilon\};\, L_{q/l\mathrm{\,Yukawa}}(y, c_N)\bigr)}
     {B\bigl(\{N\};\,  L_{q/l\mathrm{\,Yukawa}}(y, c_N)\bigr)} \ ,
\end{align}
where the numerator represents the Bayesian evidence for a given FN charge assignment $\{f, \epsilon\}$, 
and the denominator corresponds to the reference model~$N$, in which no FN symmetry is imposed.  
The likelihood $L_{q/l\mathrm{\,Yukawa}}$ is constructed solely from the Yukawa sector, 
using the observed quark (lepton) masses and mixings.  
This quantity does not depend on the parameters $m_{0}$ or $\Lambda_{B}$, as it is determined entirely by the Yukawa observables.  
Larger values of $\mathrm{BF}_{q}$ or $\mathrm{BF}_{l}$ indicate a higher degree of consistency between 
the observed Yukawa couplings and the predictions implied by the corresponding FN charge assignment.

\begin{figure}[t]
\centering
\begin{subfigure}{0.48\textwidth}
  \includegraphics[width=\linewidth]{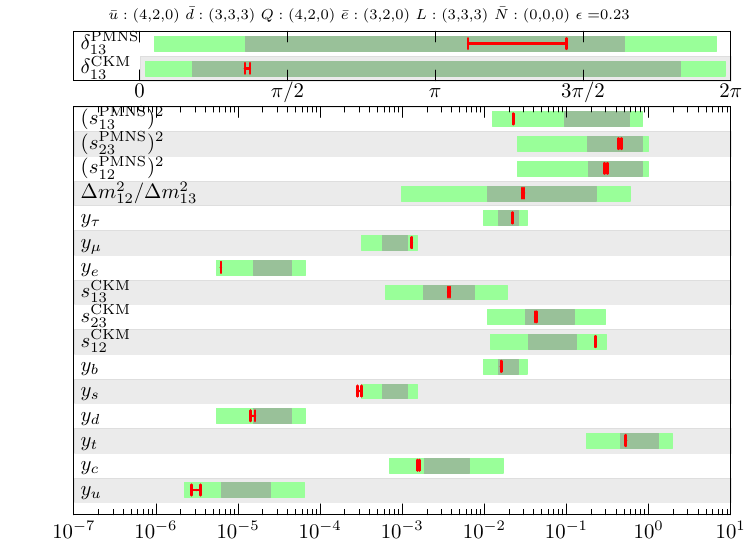}
  \caption{Benchmark $G$}
  \label{fig:Nbench}
\end{subfigure}
\hfill
\begin{subfigure}{0.48\textwidth}
  \includegraphics[width=\linewidth]{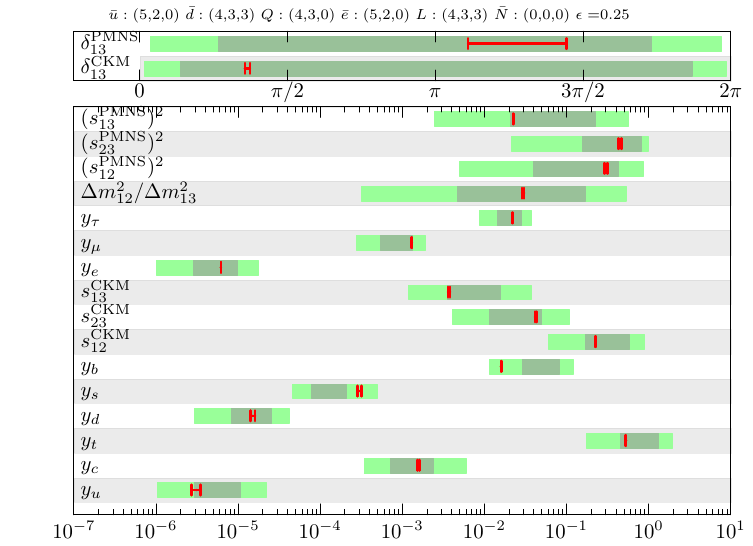}
  \caption{Benchmark $A$,$A_{\slashed{B}}$, $A'$ and $A_{H_d(3)}$}
  \label{fig:Abench}
\end{subfigure}
\begin{subfigure}{0.48\textwidth}
  \includegraphics[width=\linewidth]{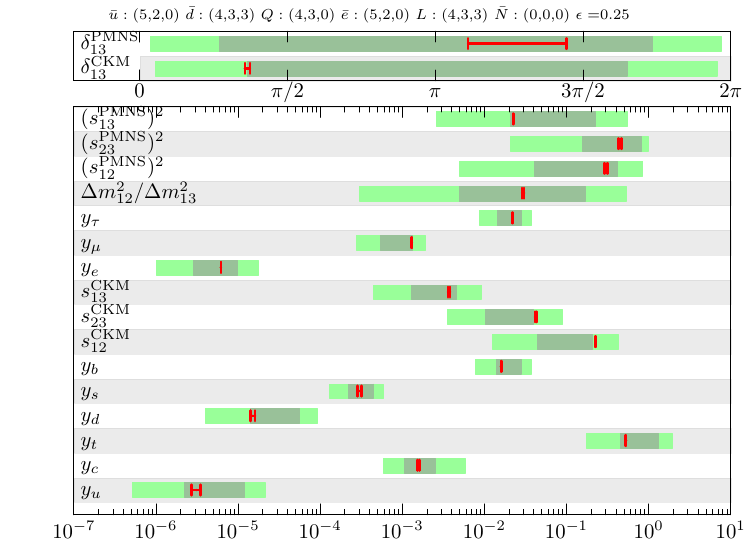}
  \caption{Benchmark $B$ and $B'$}
  \label{fig:Bbench}
\end{subfigure}
\begin{subfigure}{0.48\textwidth}
  \includegraphics[width=\linewidth]{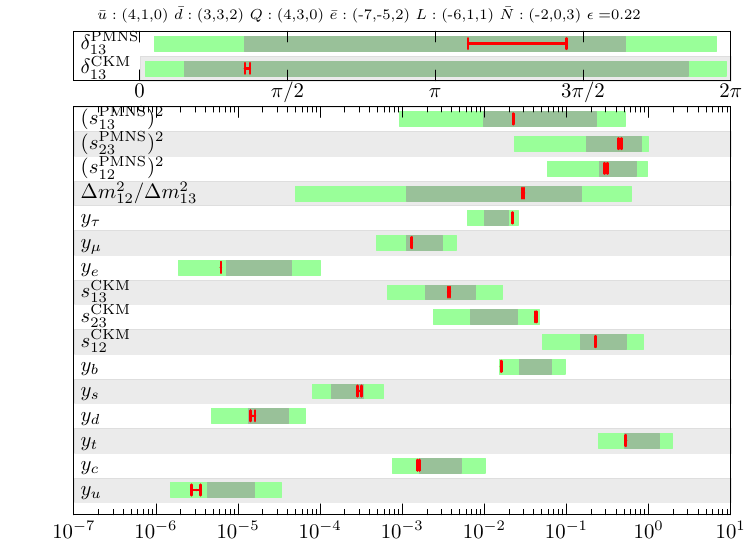}
  \caption{Benchmark $C$}
  \label{fig:Cbench}
\end{subfigure}
\caption{
Prior distributions of the Yukawa couplings and fermion mixings for representative FN charge assignments.
Model~$N$ corresponds to the no-FN scenario shown in Fig.~\ref{fig:NcaseYukawa}.
The shaded bands indicate the $1\sigma$ and $2\sigma$ ranges derived from the priors, with the experimentally measured values overlaid for comparison.
}
\label{fig:FNbenchmarks}
\end{figure}

In Fig.\,\ref{fig:FNbenchmarks}, we show the prior distributions of the fermion mixings and Yukawa couplings predicted by each benchmark model, evaluated without incorporating any experimental likelihoods. These distributions illustrate the baseline theoretical expectations associated with different FN charge assignments, allowing direct comparison among them.  
A model with a larger Bayes factor corresponds to a prior distribution that more closely reproduces the observed pattern of fermion masses and mixings, indicating a stronger intrinsic compatibility between the FN framework and experimental data.

\begin{table}[t]
\caption{
Benchmark $\mathrm{U}(1)_{\mathrm{FN}}$ charge assignments for quarks, leptons, $H_d$, and right-handed neutrinos.  
The FN charge of $H_u$ is set to zero.  
The parameter $\epsilon$ denotes the FN breaking parameter.  
Model $A_{\slashed{B}}$ has the same FN charge assignments as model $A$, but the FN mechanism is assumed not to suppress the dimension-five operators.
}
\label{tab:FNcharges}
\centering
\small
\setlength{\tabcolsep}{1.5pt}
\begin{tabular}{@{}cccccccccccc@{}}
\toprule
Benchmark
& $f_{\bar u}$
& $f_{\bar d}$
& $f_{Q}$
& $f_{\bar e}$
& $f_L$
& $f_{\bar{N}}$
& $f_{H_d}$
& $\epsilon$
& $\log_{10}\!\mathrm{BF}_{q}$
& $\log_{10}\!\mathrm{BF}_{l}$
\\
\midrule
$N$
& $(0,0,0)$
& $(0,0,0)$
& $(0,0,0)$
& $(0,0,0)$
& $(0,0,0)$
& $(0,0,0)$
& 0
& $1$
& $0$ 
& $0$ \\
$G$
& $(4,2,0)$
& $(3,3,3)$
& $(4,2,0)$
& $(4,2,0)$
& $(3,3,3)$
& $(0,0,0)$
& 0
& $0.23$
& $87$ 
& $46$ \\
$A$
& $(5,2,0)$
& $(4,3,3)$
& $(4,3,0)$
& $(5,2,0)$
& $(4,3,3)$
& $(0,0,0)$
& 0
& $0.25$
& $90$
& $47$
\\
$A_{\slashed{B}}$
& $(5,2,0)$
& $(4,3,3)$
& $(4,3,0)$
& $(5,2,0)$
& $(4,3,3)$
& $(0,0,0)$
& 0
& $0.25$
& $90$
& $47$
\\
$A'$
& $(5,2,0)$
& $(4,3,3)$
& $(4,3,0)$
& $(-5,-2,0)$
& $(-4,-3,3)$
& $(0,0,0)$
& 0
& $0.25$
& $90$
& $47$\\
$A_{H_d(3)}$
& $(5,2,0)$
& $(1,0,0)$
& $(4,3,0)$
& $(5,2,0)$
& $(1,0,0)$
& $(0,0,0)$
& 3
& $0.25$
& $90$
& $47$\\
$B$
& $(-8,2,0)$
& $(-9,3,3)$
& $(-10,3,0)$
& $(5,2,0)$
& $(4,3,3)$
& $(0,0,0)$
& 0
& $0.25$
& $89$
& $47$\\
$B'$
& $(-8,2,0)$
& $(-9,3,3)$
& $(-10,3,0)$
& $(-5,-2,0)$
& $(-4,-3,-3)$
& $(0,0,0)$
& 0
& $0.25$
& $89$
& $47$\\
$C$
& $(4,1,0)$
& $(3,3,2)$
& $(4,3,0)$
& $(-7,-5,2)$
& $(-6,1,1)$
& $(-2,0,3)$
& 0
& $0.22$
& $87$
& $46$\\
\bottomrule
\end{tabular}
\end{table}

\section{Bayesian Inference for SSM with FN Mechanism}
\label{sec:Bayesian_SSM_FN}

\subsection{Flavor/CP and Proton Decay Constraint on Sfermion Scale}

For each benchmark FN charge assignment, we evaluate the Bayes factor as a function of the sfermion mass scale $m_{0}$ and the baryon~number-violating cutoff scale $\Lambda_{B}$, and perform the Bayesian inference on $(m_{0}, \Lambda_{B})$ following the procedure described in the previous section.  
The comparison is thus carried out within the same FN setup, varying only the scales $m_{0}$ and $\Lambda_{B}$.  
The Bayes factor is normalized by the marginalized likelihood in the decoupling limit $m_{0} \to \infty$,  
\begin{align}
\mathrm{BF}(m_{0}, \Lambda_{B}, \{f, \epsilon\}; L_{\mathrm{obs}}) 
&= 
\frac{B(m_{0}, \Lambda_{B}, \{f, \epsilon\}; L_{\mathrm{obs}})}
     {B(m_{0} \to \infty, \Lambda_{B}, \{f, \epsilon\}; L_{\mathrm{obs}})} \ ,
\end{align}
where the denominator represents the reference likelihood in which all sfermions are decoupled.  
In this limit, the theory effectively reduces to the SM, and the constraints from flavor, CP, and proton decay observables no longer apply.  
If one wishes to compare models that differ in both $(m_{0}, \Lambda_{B})$ and FN charge assignments $\{f, \epsilon\}$, 
the corresponding Bayes factor can be multiplied by the Yukawa-sector Bayes factors listed in Table \ref{tab:FNcharges}, 
since the Yukawa likelihood and the low-energy likelihoods are statistically independent.

\begin{figure}[h]
\centering
\begin{subfigure}{0.45\textwidth}
  \includegraphics[width=\linewidth]{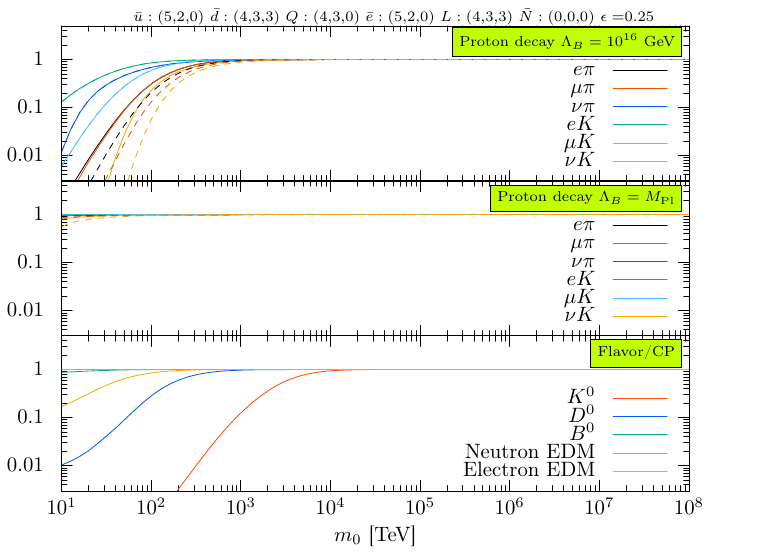}
  \caption{Benchmark $A$}
\end{subfigure}
\begin{subfigure}{0.45\textwidth}
  \includegraphics[width=\linewidth]{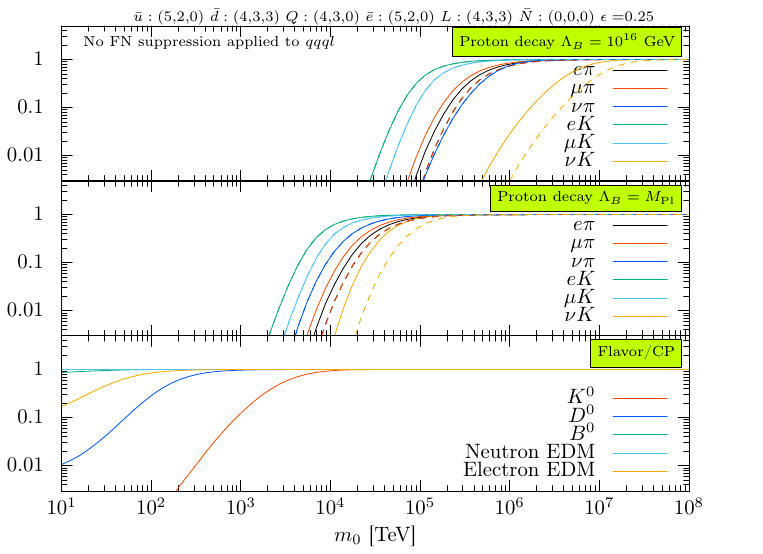}
  \caption{Benchmark $A_{\slashed{B}}$}
\end{subfigure}
\begin{subfigure}{0.45\textwidth}
  \includegraphics[width=\linewidth]{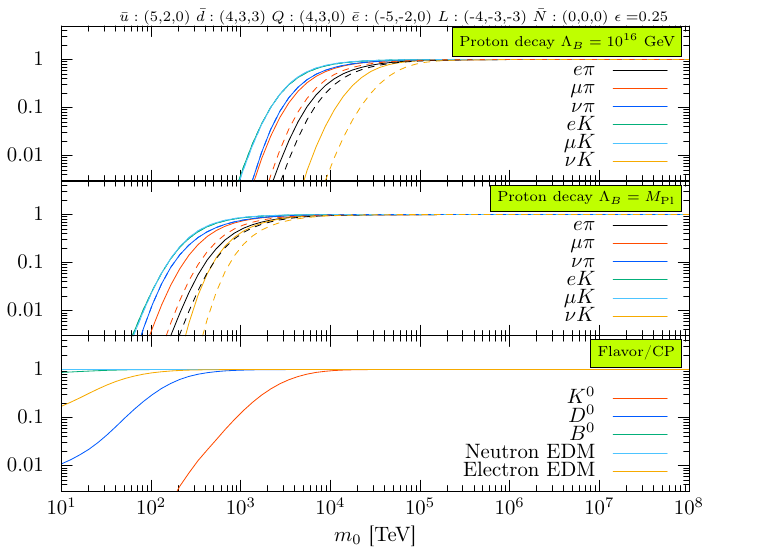}
  \caption{Benchmark $A'$}
\end{subfigure}
\begin{subfigure}{0.45\textwidth}
  \includegraphics[width=\linewidth]{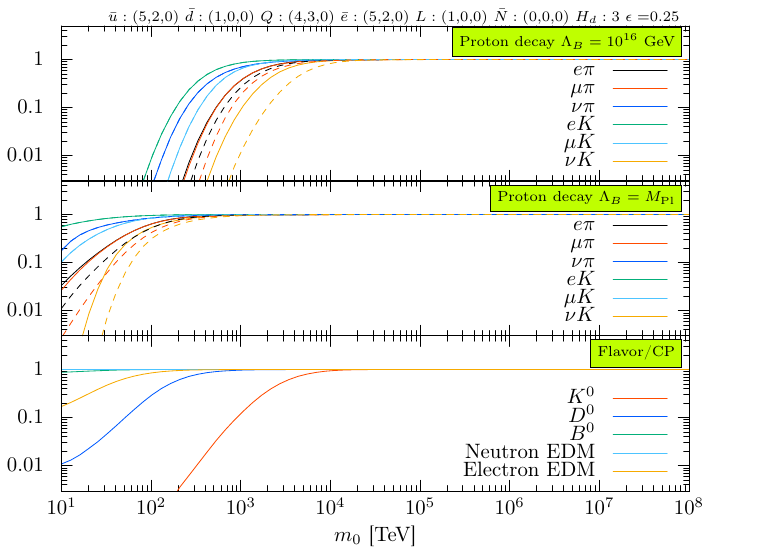}
  \caption{Benchmark $A_{H_d(3)}$}
\end{subfigure}
\caption{Bayes factors for the $A$-type benchmarks.}
\label{fig:BayesA}
\end{figure}

\begin{figure}[h]
\centering
\begin{subfigure}{0.45\textwidth}
  \includegraphics[width=\linewidth]{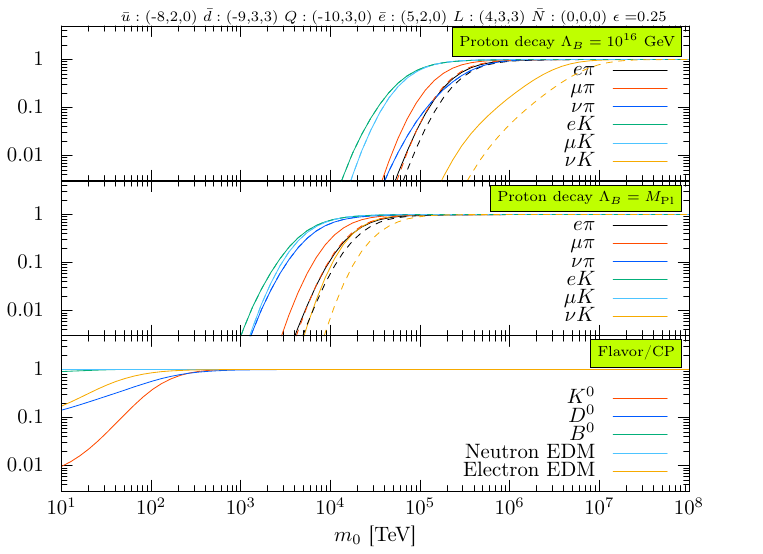}
  \caption{Benchmark $B$}
\end{subfigure}
\begin{subfigure}{0.45\textwidth}
  \includegraphics[width=\linewidth]{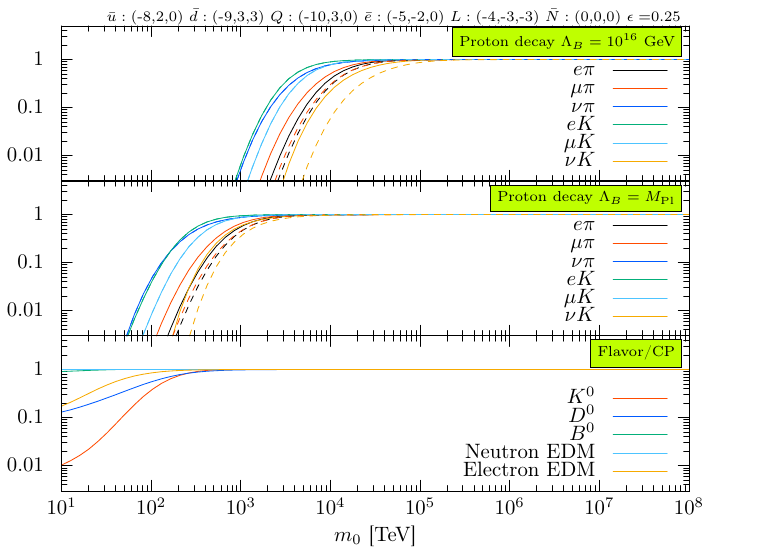}
  \caption{Benchmark $B'$}
\end{subfigure}
\caption{Bayes factors for the $B$-type benchmarks.}
\label{fig:BayesB}
\end{figure}

\begin{figure}[h]
\centering
\begin{subfigure}{0.45\textwidth}
  \includegraphics[width=\linewidth]{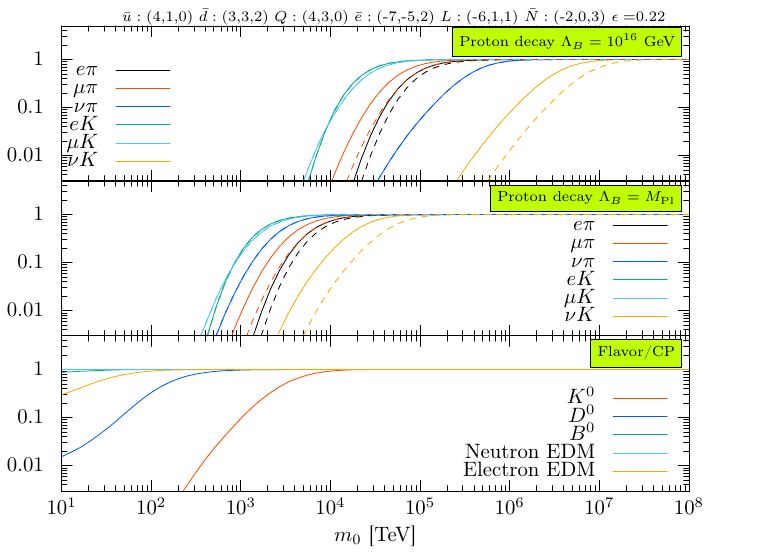}
  \caption{Benchmark $C$}
\end{subfigure}
\begin{subfigure}{0.45\textwidth}
  \includegraphics[width=\linewidth]{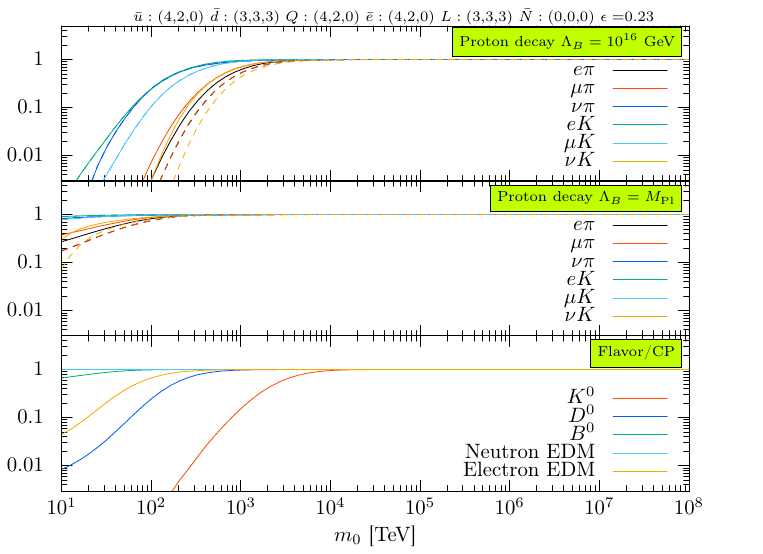}
  \caption{Benchmark $G$}
\end{subfigure}
\caption{Bayes factors for the $C$ and $G$ benchmarks.}
\label{fig:BayesCG}
\end{figure}

Figures~\ref{fig:BayesA}, \ref{fig:BayesB}, and \ref{fig:BayesCG} show the resulting Bayes factors as functions of the sfermion mass scale $m_{0}$.  
In constructing the total likelihood $L_{\mathrm{obs}}$, we include both the Yukawa-sector likelihood, which ensures consistency with the observed lepton and quark masses and mixings, and the low-energy likelihoods associated with the flavor-, CP-, and baryon~number-violating observables indicated in the figure legends.  
The combined likelihood is therefore expressed as  
\begin{align}
L_{\mathrm{obs}} = L_{\mathrm{Yukawa}} \times L_{\mathrm{FCNC/CPV/PD}} \ ,
\end{align}
where $L_{\mathrm{Yukawa}}$ encodes the fit to the measured fermion masses and mixings, while $L_{\mathrm{FCNC/CPV/PD}}$ represents the constraints from meson mixing, fermion EDMs, and proton-decay searches.  
For the proton decay contribution, we show results for two representative choices of the baryon~number-violating cutoff scale: the GUT scale, $\Lambda_{B} = 10^{16}\,\mathrm{GeV}$, and the Planck scale, $\Lambda_{B} = M_{\mathrm{Pl}}$.

Figure~\ref{fig:BayesA} shows the Bayes factors for the $A$-type charge assignments.  
Among the benchmark models considered, the $A$-type scenarios provide one of the best fits to the observed quark and lepton Yukawa couplings, as reflected in their large Yukawa-sector Bayes factors in Table~\ref{tab:FNcharges}.  
In these models, the FN charges also induce significant suppression of the dimension-five operators, resulting in a strong reduction of the proton decay rate.  
Since all $A$-type models share the same fermion and sfermion mixing structures, their Bayes factors for the flavor- and CP-violating observables are identical, and the differences in the total Bayes factors originate entirely from the proton decay sector.  

The model $A_{\slashed{B}}$ represents the case in which the FN mechanism does not work on the dimension-five operators.  
Consequently, the dimension-five operators remain unsuppressed, leading to a much faster proton decay.  
In this case, the dominant contribution arises from Higgsino dressing, whose amplitude is comparable to that in the reference model~$N$.  
The gluino-dressing contribution, on the other hand, is relatively suppressed because the sfermion flavor mixing is small in the $A$-type charge configuration.

The model $A'$ differs from $A$ only by flipping the FN charges of the lepton fields.  
Although this modification leaves the Yukawa textures unchanged, it alters the FN suppression of the baryon~number-violating dimension-five operators.  
The effective FN charges of these operators are reduced, making them less suppressed and thereby shortening the proton lifetime.  
A similar effect occurs in $A_{H_d(3)}$, where the down-type Higgs field carries a nonzero FN charge.  
In this case, the dimension-five operators are again less suppressed, leading to a proton lifetime shorter by a factor of $\epsilon^{6}$ compared with  that of the original $A$ model.

Figure~\ref{fig:BayesB} shows the Bayes factors for the $B$-type charge assignments.  
In contrast to the $A$-type models, the $B$-type benchmarks adopt a more unconventional FN charge pattern in the quark sector.  
As a result, the off-diagonal elements of the sfermion mass matrices are strongly suppressed, rendering them almost flavor diagonal.  
This structure substantially relaxes the flavor- and CP-violating constraints, so that the Bayes factors associated with the FCNC and EDM observables remain close to unity even for sfermion masses of $\mathcal{O}(100)\,\mathrm{TeV}$.  

Despite the mild flavor and CP limits, the $B$-type models are strongly constrained by proton decay bounds.  
Because the FN suppression acting on the baryon number-violating dimension-five operators is relatively weak, the predicted proton lifetime is shorter than in the conventional $A$-type model.  
Consequently, the Bayes factor decreases rapidly as $m_{0}$ is lowered, indicating that proton decay constraints dominate the overall statistical preference for these models.  

The model $B'$ differs from $B$ only by flipping the sign of the FN charges assigned to the lepton fields.  
This modification enhances the suppression of the dimension-five operators, leading to a noticeably weaker proton decay constraint than in the original $B$ model.  
Nevertheless, even in this case the predicted proton decay rate remains sizable, and the resulting Bayes factor still shows a moderate experimental tension compared with the more suppressed $A$-type scenario.

Figure~\ref{fig:BayesCG} shows the Bayes factors for the $C$- and $G$-type charge assignments.  
The $C$-type model introduces an unconventional FN charge pattern in the lepton sector.  
Although this setup can reproduce the observed Yukawa hierarchies at a reasonable level, the FN suppression of the baryon~number-violating dimension-five operators is insufficient.  
As a result, the proton decay constraint becomes particularly severe, and the Bayes factor decreases rapidly for sfermion masses of $\mathcal{O}(10)\,\mathrm{PeV}$, indicating strong experimental tension dominated by the proton decay limits.  

The $G$-type model, in contrast, closely resembles the $A$-type configuration but is motivated by an SU(5) GUT-like charge structure.  
In this case, the FN charges satisfy 
$f_{L} = f_{\bar{d}} = f_{\bar{\mathbf{5}}}$ and 
$f_{\bar{u}} = f_{Q} = f_{\bar{e}} = f_{\mathbf{10}}$, 
corresponding to the SU(5) $\bar{\mathbf{5}}$ and $\mathbf{10}$ representations, respectively.  
While this assignment yields a slightly weaker fit to the observed Yukawa structure,
which reflects the well-known tension between minimal SU(5) relations and experimental data, the resulting flavor/CP and proton decay constraints are nearly identical to those of the $A$-type model.  
Consequently, the $G$-type scenario retains a comparably large Bayes factor for sfermion masses around the PeV scale consistent with a viable supersymmetric spectrum.

\subsection{Proton Decay and Flavor Constraints at PeV Scale}

As discussed in the Introduction, PeV-scale supersymmetry remains theoretically well motivated, as it naturally accommodates the observed Higgs mass and alleviates many flavor and CP problems without resorting to excessive fine-tuning.  
Moreover, such a mass scale is also favorable in the context of anomaly mediation, which naturally leads to a TeV-scale Wino LSP consistent with the observed dark matter abundance~\cite{Hisano:2005ec}.  
In this subsection, we focus on the phenomenological implications of this regime, paying particular attention to the interplay between flavor and proton decay observables.

\begin{figure}[t]
\centering
\begin{subfigure}{0.45\textwidth}
\includegraphics[width=\linewidth]{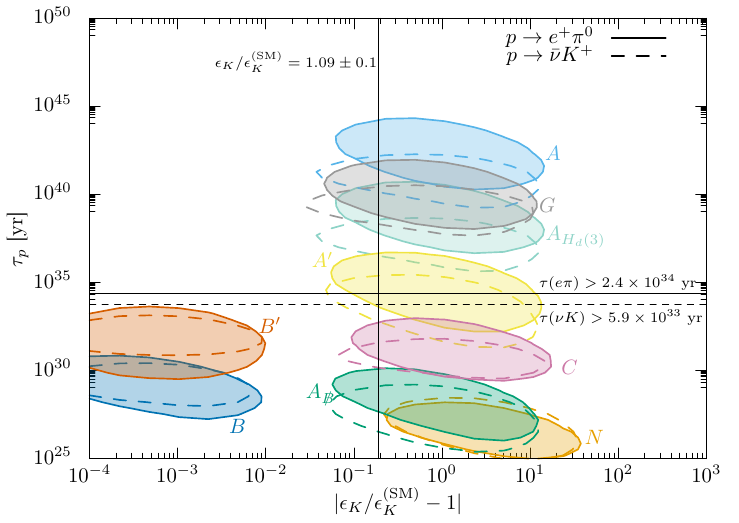}
  \caption{$1\sigma$ regions of $\epsilon_K$ and proton lifetime for each benchmark model, incorporating the Yukawa coupling constraints.}
  \label{fig:epsK_proton_2D}
\end{subfigure}
\hspace{5mm} 
\begin{subfigure}{0.45\textwidth}
\includegraphics[width=\linewidth]{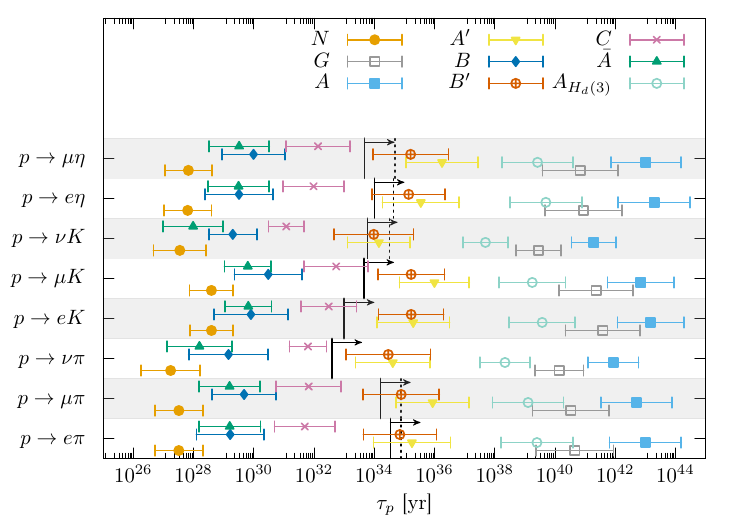}
  \caption{ $1\sigma$ regions of the predicted proton lifetimes for each decay mode, after imposing the Yukawa and flavor/CP constraints.}
  \label{fig:proton_modes}
\end{subfigure}
\caption{
Summary of Bayesian results for the benchmark models.  
Here we set $\Lambda_{B} = M_{\mathrm{Pl}}$ and $m_0=1\,\mathrm{PeV}$.
(a)~$1\sigma$ regions of $\epsilon_K$ and proton lifetime obtained after imposing the Yukawa coupling constraints.  
Solid lines correspond to the $p \to e^+\pi^0$ mode, while dashed lines indicate the $p \to \bar{\nu} K^+$ mode.  
(b)~$1\sigma$ regions of the predicted proton lifetimes for individual decay channels, incorporating the Yukawa and flavor/CP observables.
The black lines indicate the current Super-Kamiokande  limits (solid) and the projected Hyper-Kamiokande sensitivities (dashed).
}
\label{fig:oneD}
\end{figure}

Among the low-energy flavor observables, the CP-violating parameter $\epsilon_{K}$ from 
$K^0$-$\bar{K}^0$ mixing provides the most stringent constraint at present.  
We therefore examine the correlation between $\epsilon_{K}$ and the predicted proton lifetime, assuming that the baryon~number-violating cutoff scale is fixed at the Planck scale $\Lambda_{B} = M_{\mathrm{Pl}}$.  
Figure~\ref{fig:epsK_proton_2D} shows the resulting correlation between $\epsilon_{K}$ and the proton lifetime in the $p\to e^{+}\pi^{0}$ and $p\to \bar{\nu}K^{+}$ channels for each benchmark model.  
The analysis already incorporates the Yukawa coupling constraints.  
Uncertainties in the hadronic matrix elements, taken from recent lattice QCD calculations \cite{Yoo:2021gql,Aoki:2017puj}, are also included and broaden the predicted proton lifetime by approximately $50\%$.  

The behavior varies significantly among different flavor models, reflecting the underlying FN charge assignments.  
A multi-observable approach that combines information from flavor, CP, and proton decay measurements is therefore essential for disentangling the structure of the flavor symmetry underlying the theory.  

\begin{figure}[h!]
\centering
  \includegraphics[width=0.8\linewidth]{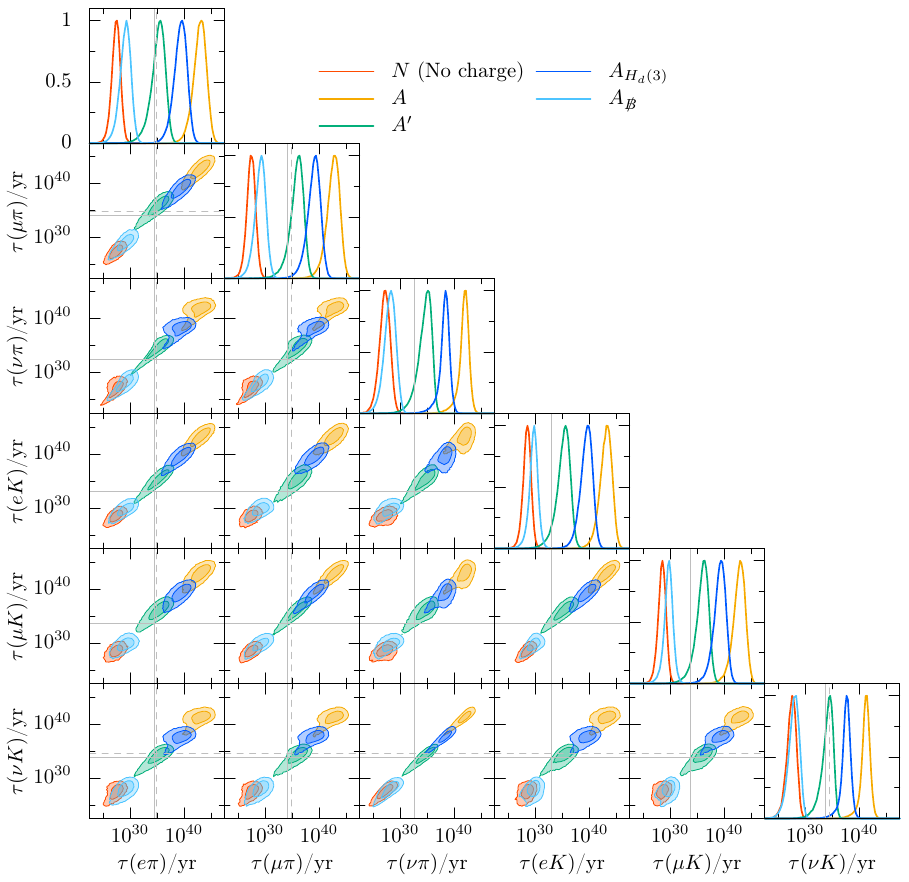}
\caption{Two-dimensional corner plots showing the $1\sigma$ regions of proton lifetimes for each decay mode in benchmark models $N$, $A$, $A'$, $A_{\slashed{B}}$, and $A_{H_d(3)}$, after imposing the Yukawa and flavor/CP constraints.
The gray lines indicate the current Super–Kamiokande  limits (solid) and the projected Hyper–Kamiokande sensitivities (dashed).
}
\label{fig:proton_A}
\end{figure}

\begin{figure}[h]
\centering
  \includegraphics[width=0.8\linewidth]{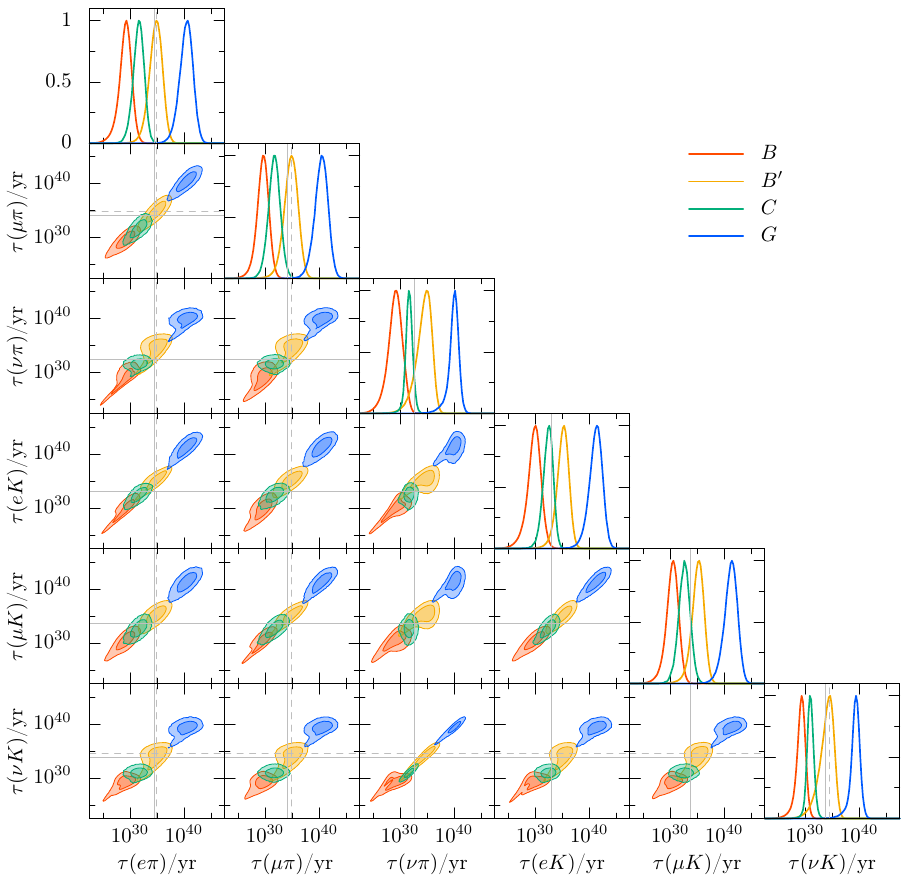}
\caption{Same as Fig.~\ref{fig:proton_A}, but for benchmark models $B$, $B'$, $C$, and $G$.
}
\label{fig:proton_B}
\end{figure}

Figure~\ref{fig:proton_modes} summarizes the predicted $1\sigma$ regions of proton lifetimes for individual decay channels after imposing both Yukawa and flavor/CP constraints.  
Furthermore, Figures~\ref{fig:proton_A} and~\ref{fig:proton_B} present detailed two-dimensional corner plots for each decay mode in the benchmark sets  
$\{N, A, A', A_{\slashed{B}}, A_{H_d(3)}\}$ and $\{B, B', C, G\}$, respectively.  
These plots display the $1\sigma$ ranges obtained after applying the full set of Yukawa and flavor/CP constraints.  
The resulting patterns of proton lifetimes differ noticeably across models and decay channels, underscoring the importance of observing multiple proton decay modes to discriminate among flavor structures in PeV-scale supersymmetry.

\subsection{Results}

In this section, we examined how different flavor models affect flavor/CP observables and nucleon decay.  
Using a Bayesian statistical framework, we quantified the consistency of each flavor model with current experimental data for given values of the sfermion mass scale $m_0$ and the baryon~number-violating cutoff scale $\Lambda_B$, and explored the resulting posterior predictions for other observables.

We find that even at the PeV scale, certain flavor models exhibit noticeable tension with precision flavor and CP measurements.  
For the representative FN charge assignments commonly used in the literature (models $A$ and $G$), the Bayes factors are typically $\mathcal{O}(0.1)$ or smaller compared to the decoupling limit $m_0 \to \infty$.  
Although not yet statistically significant, this suggests that future improvements in precision measurements and theoretical calculation could begin to probe PeV-scale supersymmetry through flavor and CP observables.

In contrast, proton decay induced by Planck-suppressed dimension-five operators is only weakly constrained for $A$ and $G$.  
Lowering the cutoff to the GUT scale, $\Lambda_B \simeq 10^{16}\,\mathrm{GeV}$, yields proton lifetimes, especially for the $p \to K\nu$ channel, within reach of future experiments such as Hyper-Kamiokande \cite{Hyper-Kamiokande:2018ofw}, JUNO \cite{JUNO:2015zny}, and DUNE \cite{DUNE:2020fgq}.  
This trend closely mirrors the predictions from colored Higgs process in minimal SU(5), since the flavor structure of the dimension-five operators is nearly identical.

The predicted lifetimes depend sensitively on the choice of flavor charge assignments.  
Models $A'$ and $A_{H_d(3)}$, for example, share the same Yukawa and sfermion structures as model $A$ but differ significantly in their dimension-five operators.  
In general, assigning zero FN charge to the Higgs fields and aligned FN charges to the SM fermions most effectively suppresses proton decay, while alternative charge configurations can shorten the lifetime, sometimes to the point of being excluded even with $\Lambda_B$ at the Planck scale.

To test the possibility that the FN symmetry does not act on Planck-suppressed operators, we also analyzed the benchmark model $A_{\slashed{B}}$.  
In this scenario, the predicted proton decay rate is substantially enhanced and already excluded experimentally.  
Although the mechanism by which global symmetries are broken by quantum gravity effects remains unclear, these results suggest that a flavor symmetry resilient to Planck-scale physics may be required.

\section{Implications from Cosmology}
\label{sec:cosmology}
In this paper, we have so far focused on low-energy observables such as flavor and CP-violating signals and proton decay constraints.  
It is also intriguing to incorporate cosmological observations into this discussion, and here we briefly outline such considerations.  
Among the cosmological aspects, the most significant is the presence of dark matter.

In the present framework, we consider a Wino LSP, as motivated by AMSB.  
The Wino mass generated through AMSB is given by
\begin{align}
    M_2|_{\mathrm{AMSB}} = \frac{g_2^2}{16\pi^2}\, m_{3/2} \ ,
\end{align}
where $g_2$ is the $\mathrm{SU}(2)_L$ gauge coupling and $m_{3/2}$ denotes the gravitino mass.  
The cosmological abundance of the Wino depends on several factors and is highly sensitive to the reheating temperature of the Universe.  
A key contribution arises from the thermal relic abundance predicted by the freeze-out mechanism, which reproduces the observed dark matter density for a Wino mass of approximately $3\,\mathrm{TeV}$.  
Therefore, if the gravitino mass is much larger than $1\,\mathrm{PeV}$, the corresponding Wino mass becomes too large, leading to an overabundance of Winos that would overclose the Universe.

At the same time, non-thermal Wino production from gravitino decays is also known to play an important role~\cite{Gherghetta:1999sw,Moroi:1999zb}.  
Several mechanisms can produce gravitinos in the early Universe.  
Since the gravitino is unstable, it eventually decays into lighter superparticles.  
The corresponding decay temperature is approximately given by
\begin{align}
    T_{\mathrm{decay}}
    \;\simeq\; 200\,\mathrm{MeV}
    \left( \frac{m_{3/2}}{1\,\mathrm{PeV}} \right)^{3/2} \ .
\end{align}
Unless the gravitino is extremely heavy, its decay takes place after Wino freeze-out in the early Universe, thereby adding a non-thermal contribution to the Wino abundance.

One of the major production channels for gravitinos in the early Universe arises from scatterings involving gauge bosons and gauginos, whose contributions scale with the reheating temperature $T_{R}$.  
The resulting non-thermal dark matter abundance from gravitino decays can be estimated as~\cite{Khlopov:1984pf,Kawasaki:1994af,Pradler:2006hh}
\begin{align}
    \Omega^{(\mathrm{reheating} \to \widetilde{G} \to \mathrm{DM})}_{\mathrm{DM}} h^2 
    \;\simeq\; 0.04
    \left( \frac{M_{\mathrm{LSP}}}{1\,\mathrm{TeV}} \right)
    \left( \frac{T_{R}}{10^{9}\,\mathrm{GeV}} \right) \, ,
\end{align}
where $M_\mathrm{LSP}$ is the mass of the LSP.
If the reheating temperature is significantly higher than $10^{9}\,\mathrm{GeV}$, the resulting dark matter abundance would exceed the observed value.

In addition, freeze-in production from sfermion decays $\widetilde{f} \to \widetilde{G}+f$ can also be significant \cite{Hall:2012zp}
if $T_R \gtrsim m_0>m_{3/2}$.  
The produced gravitinos subsequently decay, providing a non-thermal contribution to the LSP abundance given approximately by
\begin{align}
    \Omega^{(\widetilde{f} \to \widetilde{G} \to \mathrm{DM})}_{\mathrm{DM}} h^2 
    \;\sim\; 0.1 
    \left( \frac{M_\mathrm{LSP}}{1\,\mathrm{TeV}} \right)  
    \left( \frac{m_0}{10\,\mathrm{PeV}} \right)^{3}
    \left( \frac{m_{3/2}}{1\,\mathrm{PeV}} \right)^{-2} \ .
\end{align}
As a result, when the sfermion mass scale is several times larger than the gravitino mass, the total Wino abundance can become excessive, rendering such scenarios disfavored.
Consequently, taking cosmological observations into account suggests that $m_{0}\lesssim 10\,\mathrm{PeV}$ is preferred.

Furthermore, a Wino LSP dark matter candidate is subject to a variety of non-collider constraints.
In particular, indirect detection searches can provide very stringent bounds on Wino dark matter, see e.g., Refs.\,\cite{Cohen:2013ama,Fan:2013faa,Ibe:2015tma}, and in some analyses may exclude a broad mass range under certain astrophysical assumptions.
We note, however, that the quantitative strength of these bounds depends sensitively on astrophysical systematic uncertainties, such as the assumed dark-matter halo profile and background modeling in gamma-ray searches, as well as propagation/background uncertainties for cosmic-ray antiprotons (see e.g., Refs.\,\cite{Fermi-LAT:2015kyq,MAGIC:2022acl,Ibe:2015tma}).

Other cosmological probes, such as the cosmic microwave background (CMB) and big-bang nucleosynthesis (BBN), can also be relevant and are comparatively less sensitive to astrophysical modeling uncertainties.
For Wino dark matter, these considerations can translate into constraints at the level of a few hundred~GeV, e.g.\ $m_{\rm LSP}\gtrsim 300~\mathrm{GeV}$ from BBN and $m_{\rm LSP}\gtrsim 400~\mathrm{GeV}$ from the CMB (see, e.g., Refs.\,\cite{Kawasaki:2015yya,Slatyer:2015jla,Kawasaki:2021etm}).
In addition, the CMB/BBN bounds can also become relevant at higher masses where the annihilation cross section is enhanced around the Sommerfeld-resonant region near $m_{\rm LSP}\sim 2~\mathrm{TeV}$  \cite{Hisano:2003ec,Hisano:2004ds}.

A dedicated reassessment of these cosmological/astrophysical uncertainties and their impact on the viable parameter space is beyond the scope of the present paper, whose main focus is on flavor and proton-decay observables in PeV-scale SUSY. Nevertheless, incorporating these effects in a more comprehensive analysis will be an important direction for future work.

\section{Summary and Discussion}
\label{sec:conclusion}

In this work, we investigated flavor and CP observables as well as proton decay in SSMs beyond the TeV scale.  
Within this framework, various flavor models were examined to assess their consistency with current experimental data.  
We found that even for supersymmetry at the PeV scale, observable signals can still arise depending on the underlying flavor structure.  
A combined, multi-messenger approach that includes flavor, CP, and nucleon-decay observables thus provides a powerful probe of the fundamental theory, particularly its flavor symmetries.  

Assigning FN charges consistent with GUT relations can sufficiently suppress proton decay induced by Planck-suppressed operators (Model $G$).  
When the cutoff scale approaches the GUT scale, $\Lambda_B \sim 10^{16}\,\mathrm{GeV}$, the resulting proton decay rates and flavor structures become remarkably close to those predicted by SU(5) GUTs.  
While most such models remain consistent with present experimental bounds, a substantial portion of their parameter space will be probed by upcoming experiments such as Hyper-Kamiokande.  
This close correspondence between the effective FN framework and GUT expectations spotlights proton decay as a unique window into grand unification, motivating further exploration of explicit GUT realizations.  
We also found that models containing mixed-sign FN charges or Higgs fields with nonzero FN charges generally predict shorter proton lifetimes, which may be tested in the near future.  

In the present analysis, the Higgsino mass parameter has been assumed to be comparable to the sfermion mass, $|\mu| = m_{0}$.  
Relaxing this assumption can alter the relative importance of the Higgsino and gaugino dressing diagrams that mediate proton decay, depending on the origin of the $\mu$-term and on the gaugino mass spectrum.  
A substantially heavier Higgsino would
alter the gaugino spectrum expected from anomaly mediation, potentially yielding a Bino LSP (see e.g., Ref.\,\cite{Giudice:1998xp}).
In such cases, 
heavier Bino LSP potentially results in an overabundance of dark matter.  
This interplay between the Higgs sector, flavor physics, and dark matter phenomenology provides an additional motivation for precise determination of the supersymmetric mass spectrum in future studies.  

As discussed in the literature, even if the sfermion flavor structure and Yukawa couplings are governed by FN charges, Planck-suppressed dimension-five operators can still dominate if they are not subject to FN suppression.  
In such cases, the predicted proton lifetime becomes too short, requiring sfermion masses above $10^{4}\,\mathrm{TeV}$ to satisfy experimental bounds.  
Although the FN framework can partially suppress proton decay relative to anarchic flavor scenarios (Model $N$ and $A_{\slashed{B}}$), the resulting constraints remain stringent.
These results suggest that a flavor symmetry, or another symmetry such as baryon triality~\cite{Ibanez:1991pr} operative at the Planck scale may be required.

In this work, we have focused on several representative sets of FN charge assignments, but the full landscape of viable flavor structures remains to be explored.  
Moreover, our analysis assumed two chiral FN-breaking fields, $\Phi$ and $\bar{\Phi}$, with opposite charges, an assumption that may not hold in all ultraviolet completions.  
In such cases, the resulting Yukawa structures could differ substantially.  
It would also be interesting to extend the present framework to more elaborate flavor symmetries or to embed it within explicit GUT constructions, where the interplay between flavor, baryon number violation, and unification can be studied in a more complete manner.  

Finally, from a broader statistical perspective, it would be intriguing to extend the Bayesian framework employed here to include additional theoretical and cosmological inputs, such as the Higgs mass scale or the observed value of the cosmological constant.
Incorporating such quantities could offer a more unified measure of naturalness within supersymmetric theories, and would in particular require a deeper understanding of the origin of the $\mu$-term and its connection to high-scale dynamics \cite{Nomura:2014asa}.

\section*{Acknowledgements}
The authors thank Keiichi Watanabe for discussions at the early stage of this work.
This work is supported by Grant-in-Aid for Scientific Research from the Ministry of Education, Culture, Sports, Science, and Technology (MEXT), Japan, 21H04471, 22K03615, 24K23938 (M.I.), and by World Premier International Research Center Initiative (WPI), MEXT, Japan. 
This work is also supported by Grant-in-Aid for JSPS Research Fellow 24KJ0832 and by FoPM, WINGS Program, the
University of Tokyo (A.C.). 

\bibliographystyle{apsrev4-1}
\bibliography{bibtex}

@article{Ibanez:1991pr,
    author = "Ibanez, Luis E. and Ross, Graham G.",
    title = "{Discrete gauge symmetries and the origin of baryon and lepton number conservation in supersymmetric versions of the standard model}",
    reportNumber = "CERN-TH-6111-91",
    doi = "10.1016/0550-3213(92)90195-H",
    journal = "Nucl. Phys. B",
    volume = "368",
    pages = "3--37",
    year = "1992"
}

@article{Khlopov:1984pf,
    author = "Khlopov, M. Yu. and Linde, Andrei D.",
    title = "{Is It Easy to Save the Gravitino?}",
    doi = "10.1016/0370-2693(84)91656-3",
    journal = "Phys. Lett. B",
    volume = "138",
    pages = "265--268",
    year = "1984"
}

@article{Kawasaki:1994af,
    author = "Kawasaki, M. and Moroi, T.",
    title = "{Gravitino production in the inflationary universe and the effects on big bang nucleosynthesis}",
    eprint = "hep-ph/9403364",
    archivePrefix = "arXiv",
    reportNumber = "ICRR-315-94-10, TU-457",
    doi = "10.1143/PTP.93.879",
    journal = "Prog. Theor. Phys.",
    volume = "93",
    pages = "879--900",
    year = "1995"
}

@article{Moroi:1999zb,
    author = "Moroi, Takeo and Randall, Lisa",
    title = "{Wino cold dark matter from anomaly mediated SUSY breaking}",
    eprint = "hep-ph/9906527",
    archivePrefix = "arXiv",
    reportNumber = "IASSNS-HEP-99-54, PUPT-1873, MIT-CTP-2873",
    doi = "10.1016/S0550-3213(99)00748-8",
    journal = "Nucl. Phys. B",
    volume = "570",
    pages = "455--472",
    year = "2000"
}

@article{Gherghetta:1999sw,
    author = "Gherghetta, Tony and Giudice, Gian F. and Wells, James D.",
    title = "{Phenomenological consequences of supersymmetry with anomaly induced masses}",
    eprint = "hep-ph/9904378",
    archivePrefix = "arXiv",
    reportNumber = "CERN-TH-99-104",
    doi = "10.1016/S0550-3213(99)00429-0",
    journal = "Nucl. Phys. B",
    volume = "559",
    pages = "27--47",
    year = "1999"
}

@article{Hisano:2005ec,
    author = "Hisano, Junji and Matsumoto, Shigeki and Saito, Osamu and Senami, Masato",
    title = "{Heavy wino-like neutralino dark matter annihilation into antiparticles}",
    eprint = "hep-ph/0511118",
    archivePrefix = "arXiv",
    reportNumber = "ICRR-REPORT-522-2005-5, KEK-TH-1047",
    doi = "10.1103/PhysRevD.73.055004",
    journal = "Phys. Rev. D",
    volume = "73",
    pages = "055004",
    year = "2006"
}

@article{Kass:1995loi,
    author = "Kass, Robert E. and Raftery, Adrian E.",
    title = "{Bayes Factors}",
    doi = "10.1080/01621459.1995.10476572",
    journal = "J. Am. Statist. Assoc.",
    volume = "90",
    number = "430",
    pages = "773--795",
    year = "1995"
}

@book{Jeffreys:1939xee,
    author = "Jeffreys, Harold",
    title = "{The Theory of Probability}",
    isbn = "978-0-19-850368-2, 978-0-19-853193-7",
    series = "Oxford Classic Texts in the Physical Sciences",
    year = "1939"
}

@article{Weinberg:1979sa,
    author = "Weinberg, Steven",
    title = "{Baryon and Lepton Nonconserving Processes}",
    reportNumber = "HUTP-79-A050",
    doi = "10.1103/PhysRevLett.43.1566",
    journal = "Phys. Rev. Lett.",
    volume = "43",
    pages = "1566--1570",
    year = "1979"
}

@article{Carone:1996nd,
    author = "Carone, Christopher D. and Hall, Lawrence J. and Murayama, Hitoshi",
    title = "{A Supersymmetric theory of flavor and R-parity}",
    eprint = "hep-ph/9602364",
    archivePrefix = "arXiv",
    reportNumber = "LBL-38380, UCB-PTH-96-06",
    doi = "10.1103/PhysRevD.54.2328",
    journal = "Phys. Rev. D",
    volume = "54",
    pages = "2328--2339",
    year = "1996"
}

@article{Carone:1995xw,
    author = "Carone, Christopher D. and Hall, Lawrence J. and Murayama, Hitoshi",
    title = "{S(3)3 flavor symmetry and p ---{\ensuremath{>}} K0 e+}",
    eprint = "hep-ph/9512399",
    archivePrefix = "arXiv",
    reportNumber = "LBL-38047, UCB-PTH-95-43",
    doi = "10.1103/PhysRevD.53.6282",
    journal = "Phys. Rev. D",
    volume = "53",
    pages = "6282--6291",
    year = "1996"
}

@article{Kakizaki:2002hs,
    author = "Kakizaki, Mitsuru and Yamaguchi, Masahiro",
    title = "{U(1) flavor symmetry and proton decay in supersymmetric standard model}",
    eprint = "hep-ph/0203192",
    archivePrefix = "arXiv",
    reportNumber = "TU-648",
    doi = "10.1088/1126-6708/2002/06/032",
    journal = "JHEP",
    volume = "06",
    pages = "032",
    year = "2002"
}

@article{Nagata:2013sba,
    author = "Nagata, Natsumi and Shirai, Satoshi",
    title = "{Sfermion Flavor and Proton Decay in High-Scale Supersymmetry}",
    eprint = "1312.7854",
    archivePrefix = "arXiv",
    primaryClass = "hep-ph",
    doi = "10.1007/JHEP03(2014)049",
    journal = "JHEP",
    volume = "03",
    pages = "049",
    year = "2014"
}

@article{Csaki:2013we,
    author = "Csaki, Csaba and Heidenreich, Ben",
    title = "{A Complete Model for R-parity Violation}",
    eprint = "1302.0004",
    archivePrefix = "arXiv",
    primaryClass = "hep-ph",
    doi = "10.1103/PhysRevD.88.055023",
    journal = "Phys. Rev. D",
    volume = "88",
    pages = "055023",
    year = "2013"
}

@article{Nelson:1997bt,
    author = "Nelson, Ann E. and Wright, David",
    title = "{Horizontal, anomalous U(1) symmetry for the more minimal supersymmetric standard model}",
    eprint = "hep-ph/9702359",
    archivePrefix = "arXiv",
    reportNumber = "UW-PT-97-03",
    doi = "10.1103/PhysRevD.56.1598",
    journal = "Phys. Rev. D",
    volume = "56",
    pages = "1598--1604",
    year = "1997"
}

@article{Ben-Hamo:1994dha,
    author = "Ben-Hamo, Valerie and Nir, Yosef",
    title = "{Implications of horizontal symmetries on baryon number violation in supersymmetric models}",
    eprint = "hep-ph/9408315",
    archivePrefix = "arXiv",
    reportNumber = "WIS-94-34-PH",
    doi = "10.1016/0370-2693(94)91135-5",
    journal = "Phys. Lett. B",
    volume = "339",
    pages = "77--82",
    year = "1994"
}

@article{Murayama:1994tc,
    author = "Murayama, Hitoshi and Kaplan, D. B.",
    title = "{Family symmetries and proton decay}",
    eprint = "hep-ph/9406423",
    archivePrefix = "arXiv",
    reportNumber = "NSF-ITP-94-69, LBL-35807, DOE-ER-40561-148, INT-94-00-61",
    doi = "10.1016/0370-2693(94)90242-9",
    journal = "Phys. Lett. B",
    volume = "336",
    pages = "221--228",
    year = "1994"
}

@article{Dine:2013nga,
    author = "Dine, Michael and Draper, Patrick and Shepherd, William",
    title = "{Proton decay at $M_{pl}$ and the scale of SUSY-breaking}",
    eprint = "1308.0274",
    archivePrefix = "arXiv",
    primaryClass = "hep-ph",
    reportNumber = "SCIPP-13-08",
    doi = "10.1007/JHEP02(2014)027",
    journal = "JHEP",
    volume = "02",
    pages = "027",
    year = "2014"
}

@article{Sakai:1981pk,
    author = "Sakai, N. and Yanagida, Tsutomu",
    title = "{Proton Decay in a Class of Supersymmetric Grand Unified Models}",
    reportNumber = "MPI-PAE/PTh 55/81",
    doi = "10.1016/0550-3213(82)90457-6",
    journal = "Nucl. Phys. B",
    volume = "197",
    pages = "533",
    year = "1982"
}

@article{Dimopoulos:1981zb,
    author = "Dimopoulos, Savas and Georgi, Howard",
    title = "{Softly Broken Supersymmetry and SU(5)}",
    reportNumber = "HUTP-81/A022",
    doi = "10.1016/0550-3213(81)90522-8",
    journal = "Nucl. Phys. B",
    volume = "193",
    pages = "150--162",
    year = "1981"
}

@article{Sakai:1981gr,
    author = "Sakai, N.",
    title = "{Naturalness in Supersymmetric Guts}",
    reportNumber = "TU/81/225",
    doi = "10.1007/BF01573998",
    journal = "Z. Phys. C",
    volume = "11",
    pages = "153",
    year = "1981"
}

@inproceedings{Maiani:1979gif,
  author       = {Luciano Maiani},
  title        = {Vector bosons and Higgs bosons in the Salam-Weinberg theory of weak and electromagnetic interactions},
  booktitle    = {Proceedings of the 11th GIF Summer School on Particle Physics},
  editor       = {M. Davier and others},
  address      = {Gif-sur-Yvette, France},
  publisher    = {IN2P3},
  year         = {1980},
  pages        = {1--52},
  note         = {Presented at Gif-sur-Yvette, 3--7 September 1979}
}

@article{Gabbiani:1996hi,
    author = "Gabbiani, F. and Gabrielli, E. and Masiero, A. and Silvestrini, L.",
    title = "{A Complete analysis of FCNC and CP constraints in general SUSY extensions of the standard model}",
    eprint = "hep-ph/9604387",
    archivePrefix = "arXiv",
    reportNumber = "ROM2F-96-21",
    doi = "10.1016/0550-3213(96)00390-2",
    journal = "Nucl. Phys. B",
    volume = "477",
    pages = "321--352",
    year = "1996"
}

@article{Wells:2004di,
    author = "Wells, James D.",
    title = "{PeV-scale supersymmetry}",
    eprint = "hep-ph/0411041",
    archivePrefix = "arXiv",
    reportNumber = "MCTP-04-61",
    doi = "10.1103/PhysRevD.71.015013",
    journal = "Phys. Rev. D",
    volume = "71",
    pages = "015013",
    year = "2005"
}

@article{Arkani-Hamed:2004zhs,
    author = "Arkani-Hamed, N. and Dimopoulos, S. and Giudice, G. F. and Romanino, A.",
    title = "{Aspects of split supersymmetry}",
    eprint = "hep-ph/0409232",
    archivePrefix = "arXiv",
    reportNumber = "CERN-PH-TH-2004-183",
    doi = "10.1016/j.nuclphysb.2004.12.026",
    journal = "Nucl. Phys. B",
    volume = "709",
    pages = "3--46",
    year = "2005"
}

@article{Giudice:2004tc,
    author = "Giudice, G. F. and Romanino, A.",
    title = "{Split supersymmetry}",
    eprint = "hep-ph/0406088",
    archivePrefix = "arXiv",
    reportNumber = "CERN-PH-TH-2004-100",
    doi = "10.1016/j.nuclphysb.2004.08.001",
    journal = "Nucl. Phys. B",
    volume = "699",
    pages = "65--89",
    year = "2004",
    note = "[Erratum: Nucl.Phys.B 706, 487--487 (2005)]"
}

@article{Arkani-Hamed:2004ymt,
    author = "Arkani-Hamed, Nima and Dimopoulos, Savas",
    title = "{Supersymmetric unification without low energy supersymmetry and signatures for fine-tuning at the LHC}",
    eprint = "hep-th/0405159",
    archivePrefix = "arXiv",
    doi = "10.1088/1126-6708/2005/06/073",
    journal = "JHEP",
    volume = "06",
    pages = "073",
    year = "2005"
}

@inproceedings{Wells:2003tf,
    author = "Wells, James D.",
    title = "{Implications of supersymmetry breaking with a little hierarchy between gauginos and scalars}",
    booktitle = "{11th International Conference on Supersymmetry and the Unification of Fundamental Interactions}",
    eprint = "hep-ph/0306127",
    archivePrefix = "arXiv",
    reportNumber = "MCTP-03-30",
    month = "6",
    year = "2003"
}

@article{Okada:1990vk,
    author = "Okada, Yasuhiro and Yamaguchi, Masahiro and Yanagida, Tsutomu",
    title = "{Renormalization Group Analysis on the Higgs Mass in the Softly Broken Supersymmetric Standard Model}",
    journal = "Prog. Theor. Phys.",
    volume = "85",
    pages = "1-6",
    year = "1991",
    doi = "10.1143/ptp/85.1.1"
}

@article{Okada:1991jc,
    author = "Okada, Yasuhiro and Yamaguchi, Masahiro and Yanagida, Tsutomu",
    title = "{Upper Bound of the Lightest Higgs Boson Mass in the Minimal Supersymmetric Standard Model}",
    journal = "Phys. Lett. B",
    volume = "262",
    pages = "54-58",
    year = "1991",
    doi = "10.1016/0370-2693(91)90642-4"
}

@article{Ellis:1990nz,
    author = "Ellis, John R. and Ridolfi, Giovanni and Zwirner, Fabio",
    title = "{Radiative Corrections to the Masses of Supersymmetric Higgs Bosons}",
    journal = "Phys. Lett. B",
    volume = "257",
    pages = "83-91",
    year = "1991",
    doi = "10.1016/0370-2693(91)90863-L"
}

@article{Haber:1990aw,
    author = "Haber, Howard E. and Hempfling, R.",
    title = "{Can the Mass of the Lightest Higgs Boson of the Minimal Supersymmetric Model Be Larger Than m(Z)?}",
    journal = "Phys. Rev. Lett.",
    volume = "66",
    pages = "1815-1818",
    year = "1991",
    doi = "10.1103/PhysRevLett.66.1815"
}

@article{Ellis:1991zd,
    author = "Ellis, John R. and Ridolfi, Giovanni and Zwirner, Fabio",
    title = "{On radiative corrections to supersymmetric Higgs boson masses and their implications for LEP searches}",
    journal = "Phys. Lett. B",
    volume = "262",
    pages = "477-484",
    year = "1991",
    doi = "10.1016/0370-2693(91)90626-2"
}

@article{Hall:2011jd,
    author = "Hall, Lawrence J. and Nomura, Yasunori",
    title = "{Spread Supersymmetry}",
    eprint = "1111.4519",
    archivePrefix = "arXiv",
    primaryClass = "hep-ph",
    reportNumber = "UCB-PTH-11-09",
    doi = "10.1007/JHEP01(2012)082",
    journal = "JHEP",
    volume = "01",
    pages = "082",
    year = "2012"
}

@article{Hall:2012zp,
    author = "Hall, Lawrence J. and Nomura, Yasunori and Shirai, Satoshi",
    title = "{Spread Supersymmetry with Wino LSP: Gluino and Dark Matter Signals}",
    eprint = "1210.2395",
    archivePrefix = "arXiv",
    primaryClass = "hep-ph",
    reportNumber = "MIT-CTP-4402, UCB-PTH-12-16",
    doi = "10.1007/JHEP01(2013)036",
    journal = "JHEP",
    volume = "01",
    pages = "036",
    year = "2013"
}

@article{Nomura:2014asa,
    author = "Nomura, Yasunori and Shirai, Satoshi",
    title = "{Supersymmetry from Typicality: TeV-Scale Gauginos and PeV-Scale Squarks and Sleptons}",
    eprint = "1407.3785",
    archivePrefix = "arXiv",
    primaryClass = "hep-ph",
    reportNumber = "UCB-PTH-14-32",
    doi = "10.1103/PhysRevLett.113.111801",
    journal = "Phys. Rev. Lett.",
    volume = "113",
    number = "11",
    pages = "111801",
    year = "2014"
}

@article{Ibe:2012hu,
    author = "Ibe, Masahiro and Matsumoto, Shigeki and Yanagida, Tsutomu T.",
    title = "{Pure Gravity Mediation with $m_{3/2} = 10$--$100$ TeV}",
    eprint = "1202.2253",
    archivePrefix = "arXiv",
    primaryClass = "hep-ph",
    reportNumber = "ICRR-REPORT-605-2011-22, IPMU11-0016",
    doi = "10.1103/PhysRevD.85.095011",
    journal = "Phys. Rev. D",
    volume = "85",
    pages = "095011",
    year = "2012"
}

@article{Arvanitaki:2012ps,
    author = "Arvanitaki, Asimina and Craig, Nathaniel and Dimopoulos, Savas and Villadoro, Giovanni",
    title = "{Mini-Split}",
    eprint = "1210.0555",
    archivePrefix = "arXiv",
    primaryClass = "hep-ph",
    doi = "10.1007/JHEP02(2013)126",
    journal = "JHEP",
    volume = "02",
    pages = "126",
    year = "2013"
}

@article{ArkaniHamed:2012gw,
    author = "Arkani-Hamed, Nima and Gupta, Arpit and Kaplan, David E. and Weiner, Neal and Zorawski, Tom",
    title = "{Simply Unnatural Supersymmetry}",
    eprint = "1212.6971",
    journal = "$\phantom{X}$",
    archivePrefix = "arXiv",
    primaryClass = "hep-ph",
    month = "12",
    year = "2012"
}

@article{Ibe:2011aa,
    author = "Ibe, Masahiro and Yanagida, Tsutomu T.",
    title = "{The Lightest Higgs Boson Mass in Pure Gravity Mediation Model}",
    eprint = "1112.2462",
    archivePrefix = "arXiv",
    primaryClass = "hep-ph",
    reportNumber = "IPMU11-0207, ICRR-REPORT-601-2011-18",
    doi = "10.1016/j.physletb.2012.02.034",
    journal = "Phys. Lett. B",
    volume = "709",
    pages = "374--380",
    year = "2012"
}

@article{Randall:1998uk,
    author = "Randall, Lisa and Sundrum, Raman",
    title = "{Out of this world supersymmetry breaking}",
    eprint = "hep-th/9810155",
    archivePrefix = "arXiv",
    reportNumber = "MIT-CTP-2788, PUPT-1815, BUHEP-98-26",
    doi = "10.1016/S0550-3213(99)00359-4",
    journal = "Nucl. Phys. B",
    volume = "557",
    pages = "79--118",
    year = "1999"
}

@article{Giudice:1998xp,
    author = "Giudice, Gian F. and Luty, Markus A. and Murayama, Hitoshi and Rattazzi, Riccardo",
    title = "{Gaugino mass without singlets}",
    eprint = "hep-ph/9810442",
    archivePrefix = "arXiv",
    reportNumber = "CERN-TH-98-337, LBNL-42419, LBL-42419, UCB-PTH-98-50, UMD-PP-99-037",
    doi = "10.1088/1126-6708/1998/12/027",
    journal = "JHEP",
    volume = "12",
    pages = "027",
    year = "1998"
}

@article{Minkowski:1977sc,
    author = "Minkowski, Peter",
    title = "{$\mu \to e\gamma$ at a Rate of One Out of $10^{9}$ Muon Decays?}",
    reportNumber = "Print-77-0182 (BERN)",
    doi = "10.1016/0370-2693(77)90435-X",
    journal = "Phys. Lett. B",
    volume = "67",
    pages = "421--428",
    year = "1977"
}

@article{Yanagida:1979as,
    author = "Yanagida, Tsutomu",
    editor = "Sawada, Osamu and Sugamoto, Akio",
    title = "{Horizontal gauge symmetry and masses of neutrinos}",
    reportNumber = "KEK-79-18-95",
    journal = "Conf. Proc. C",
    volume = "7902131",
    pages = "95--99",
    year = "1979"
}

@article{Yanagida:1979gs,
    author = "Yanagida, Tsutomu",
    title = "{Horizontal Symmetry and Mass of the Top Quark}",
    reportNumber = "TU/79/196",
    doi = "10.1103/PhysRevD.20.2986",
    journal = "Phys. Rev. D",
    volume = "20",
    pages = "2986",
    year = "1979"
}

@article{Gell-Mann:1979vob,
    author = "Gell-Mann, Murray and Ramond, Pierre and Slansky, Richard",
    title = "{Complex Spinors and Unified Theories}",
    eprint = "1306.4669",
    archivePrefix = "arXiv",
    primaryClass = "hep-th",
    reportNumber = "PRINT-80-0576",
    journal = "Conf. Proc. C",
    volume = "790927",
    pages = "315--321",
    year = "1979"
}

@article{Glashow:1979nm,
    author = "Glashow, S. L.",
    editor = "L{\'e}vy, Maurice and Basdevant, Jean-Louis and Speiser, David and Weyers, Jacques and Gastmans, Raymond and Jacob, Maurice",
    title = "{The Future of Elementary Particle Physics}",
    reportNumber = "HUTP-79-A059",
    doi = "10.1007/978-1-4684-7197-7_15",
    journal = "NATO Sci. Ser. B",
    volume = "61",
    pages = "687",
    year = "1980"
}

@article{Mohapatra:1979ia,
    author = "Mohapatra, Rabindra N. and Senjanovic, Goran",
    title = "{Neutrino Mass and Spontaneous Parity Nonconservation}",
    reportNumber = "MDDP-TR-80-060, MDDP-PP-80-105, CCNY-HEP-79-10",
    doi = "10.1103/PhysRevLett.44.912",
    journal = "Phys. Rev. Lett.",
    volume = "44",
    pages = "912",
    year = "1980"
}

@article{2201.02472,
    author = "Aad, Georges and others",
    collaboration = "ATLAS",
    title = "{Search for long-lived charginos based on a disappearing-track signature using 136 fb$^{-1}$ of pp collisions at $\sqrt{s}$~=~13~TeV with the ATLAS detector}",
    eprint = "2201.02472",
    archivePrefix = "arXiv",
    primaryClass = "hep-ex",
    reportNumber = "CERN-EP-2021-209",
    doi = "10.1140/epjc/s10052-022-10489-5",
    journal = "Eur. Phys. J. C",
    volume = "82",
    number = "7",
    pages = "606",
    year = "2022"
}

@article{2309.16823,
    author = "Hayrapetyan, Aram and others",
    collaboration = "CMS",
    title = "{Search for supersymmetry in final states with disappearing tracks in proton-proton collisions at s=13{\,}{\,}TeV}",
    eprint = "2309.16823",
    archivePrefix = "arXiv",
    primaryClass = "hep-ex",
    reportNumber = "CMS-SUS-21-006, CERN-EP-2023-209",
    doi = "10.1103/PhysRevD.109.072007",
    journal = "Phys. Rev. D",
    volume = "109",
    number = "7",
    pages = "072007",
    year = "2024"
}

@article{2210.16035,
    author = "Ibe, Masahiro and Mishima, Masataka and Nakayama, Yuhei and Shirai, Satoshi",
    title = "{Precise estimate of charged Wino decay rate}",
    eprint = "2210.16035",
    archivePrefix = "arXiv",
    primaryClass = "hep-ph",
    reportNumber = "IPMU22-0054",
    doi = "10.1007/JHEP01(2023)017",
    journal = "JHEP",
    volume = "01",
    pages = "017",
    year = "2023"
}

@article{2312.08087,
    author = "Ibe, Masahiro and Nakayama, Yuhei and Shirai, Satoshi",
    title = "{Precise estimate of charged Higgsino/Wino decay rate}",
    eprint = "2312.08087",
    archivePrefix = "arXiv",
    primaryClass = "hep-ph",
    reportNumber = "IPMU23-0048",
    doi = "10.1007/JHEP03(2024)012",
    journal = "JHEP",
    volume = "03",
    pages = "012",
    year = "2024"
}

@article{2404.02297,
    author = "Boyle, Peter A. and Erben, Felix and Flynn, Jonathan M. and Garron, Nicolas and Kettle, Julia and Mukherjee, Rajnandini and Tsang, J. Tobias",
    collaboration = "RBC, UKQCD",
    title = "{Kaon mixing beyond the standard model with physical masses}",
    eprint = "2404.02297",
    archivePrefix = "arXiv",
    primaryClass = "hep-lat",
    reportNumber = "CERN-TH-2024-040, LTH 1366",
    doi = "10.1103/PhysRevD.110.034501",
    journal = "Phys. Rev. D",
    volume = "110",
    number = "3",
    pages = "034501",
    year = "2024"
}

@article{1907.01025,
    author = "Dowdall, R. J. and Davies, C. T. H. and Horgan, R. R. and Lepage, G. P. and Monahan, C. J. and Shigemitsu, J. and Wingate, M.",
    title = "{Neutral B-meson mixing from full lattice QCD at the physical point}",
    eprint = "1907.01025",
    archivePrefix = "arXiv",
    primaryClass = "hep-lat",
    reportNumber = "INT-PUB-19-031, JLAB-THY-19-3068",
    doi = "10.1103/PhysRevD.100.094508",
    journal = "Phys. Rev. D",
    volume = "100",
    number = "9",
    pages = "094508",
    year = "2019"
}

@article{1706.04622,
    author = "Bazavov, A. and others",
    title = "{Short-distance matrix elements for $D^0$-meson mixing for $N_f=2+1$ lattice QCD}",
    eprint = "1706.04622",
    archivePrefix = "arXiv",
    primaryClass = "hep-lat",
    reportNumber = "FERMILAB-PUB-17-196-T",
    doi = "10.1103/PhysRevD.97.034513",
    journal = "Phys. Rev. D",
    volume = "97",
    number = "3",
    pages = "034513",
    year = "2018"
}

@article{2411.18639,
    author = "Banerjee, Swagato and others",
    collaboration = "Heavy Flavor Averaging Group (HFLAV)",
    title = "{Averages of $b$-hadron, $c$-hadron, and $\tau$-lepton properties as of 2023}",
    eprint = "2411.18639",
    archivePrefix = "arXiv",
    primaryClass = "hep-ex",
    month = "11",
    year = "2024"
}

@article{2212.11841,
    author = "Roussy, Tanya S. and others",
    title = "{An improved bound on the electron{\textquoteright}s electric dipole moment}",
    eprint = "2212.11841",
    archivePrefix = "arXiv",
    primaryClass = "physics.atom-ph",
    doi = "10.1126/science.adg4084",
    journal = "Science",
    volume = "381",
    number = "6653",
    pages = "adg4084",
    year = "2023"
}

@article{2303.02822,
    author = "Kaneta, Kunio and Nagata, Natsumi and Olive, Keith A. and Pospelov, Maxim and Velasco-Sevilla, Liliana",
    title = "{Quantifying limits on CP violating phases from EDMs in supersymmetry}",
    eprint = "2303.02822",
    archivePrefix = "arXiv",
    primaryClass = "hep-ph",
    reportNumber = "UMN-TH-4208/23, FTPI-MINN-23/02, CQUeST-2023-0719",
    doi = "10.1007/JHEP03(2023)250",
    journal = "JHEP",
    volume = "03",
    pages = "250",
    year = "2023"
}

@article{2412.19484,
    author = "Ibe, Masahiro and Shirai, Satoshi and Watanabe, Keiichi",
    title = "{Comprehensive Bayesian exploration of Froggatt-Nielsen mechanism}",
    eprint = "2412.19484",
    archivePrefix = "arXiv",
    primaryClass = "hep-ph",
    reportNumber = "IPMU24-0047",
    doi = "10.1007/JHEP03(2025)150",
    journal = "JHEP",
    volume = "03",
    pages = "150",
    year = "2025"
}

@article{Weinberg:1981wj,
    author = "Weinberg, Steven",
    title = "{Supersymmetry at Ordinary Energies. 1. Masses and Conservation Laws}",
    reportNumber = "HUTP-81-A047",
    doi = "10.1103/PhysRevD.26.287",
    journal = "Phys. Rev. D",
    volume = "26",
    pages = "287",
    year = "1982"
}

@article{UTfit:2022hsi,
    author = "Bona, Marcella and others",
    collaboration = "UTfit",
    title = "{New UTfit Analysis of the Unitarity Triangle in the Cabibbo-Kobayashi-Maskawa scheme}",
    eprint = "2212.03894",
    archivePrefix = "arXiv",
    primaryClass = "hep-ph",
    reportNumber = "YITP-SB-2022-40",
    doi = "10.1007/s12210-023-01137-5",
    journal = "Rend. Lincei Sci. Fis. Nat.",
    volume = "34",
    pages = "37--57",
    year = "2023",
    note = "Updated results available at \url{http://www.utfit.org/UTfit/ResultsSummer2023NP}"
}

@article{Belyaev:1982ik,
    author = "Belyaev, V. M. and Vysotsky, M. I.",
    title = "{MORE ABOUT PROTON DECAY DUE TO d = 5 OPERATORS}",
    reportNumber = "ITEP-163-1982",
    doi = "10.1016/0370-2693(83)90879-1",
    journal = "Phys. Lett. B",
    volume = "127",
    pages = "215--218",
    year = "1983"
}

@article{Altmannshofer:2013lfa,
    author = "Altmannshofer, Wolfgang and Harnik, Roni and Zupan, Jure",
    title = "{Low Energy Probes of PeV Scale Sfermions}",
    eprint = "1308.3653",
    archivePrefix = "arXiv",
    primaryClass = "hep-ph",
    reportNumber = "FERMILAB-PUB-13-319-T",
    doi = "10.1007/JHEP11(2013)202",
    journal = "JHEP",
    volume = "11",
    pages = "202",
    year = "2013"
}

@article{McKeen:2013dma,
    author = "McKeen, David and Pospelov, Maxim and Ritz, Adam",
    title = "{Electric dipole moment signatures of PeV-scale superpartners}",
    eprint = "1303.1172",
    archivePrefix = "arXiv",
    primaryClass = "hep-ph",
    doi = "10.1103/PhysRevD.87.113002",
    journal = "Phys. Rev. D",
    volume = "87",
    number = "11",
    pages = "113002",
    year = "2013"
}

@article{Abel:2020pzs,
    author = "Abel, C. and others",
    title = "{Measurement of the Permanent Electric Dipole Moment of the Neutron}",
    eprint = "2001.11966",
    archivePrefix = "arXiv",
    primaryClass = "hep-ex",
    doi = "10.1103/PhysRevLett.124.081803",
    journal = "Phys. Rev. Lett.",
    volume = "124",
    number = "8",
    pages = "081803",
    year = "2020"
}

@article{ParticleDataGroup:2022pth,
    author = "Workman, R. L. and others",
    collaboration = "Particle Data Group",
    title = "{Review of Particle Physics}",
    doi = "10.1093/ptep/ptac097",
    journal = "PTEP",
    volume = "2022",
    pages = "083C01",
    year = "2022"
}

@article{Buttazzo:2013uya,
    author = "Buttazzo, Dario and Degrassi, Giuseppe and Giardino, Pier Paolo and Giudice, Gian F. and Sala, Filippo and Salvio, Alberto and Strumia, Alessandro",
    title = "{Investigating the near-criticality of the Higgs boson}",
    eprint = "1307.3536",
    archivePrefix = "arXiv",
    primaryClass = "hep-ph",
    reportNumber = "CERN-PH-TH-2013-166, FTUAM-13-20, IFT-UAM-CSIC-13-081, IFUP-TH",
    doi = "10.1007/JHEP12(2013)089",
    journal = "JHEP",
    volume = "12",
    pages = "089",
    year = "2013"
}

@article{Chetyrkin:1997sg,
    author = "Chetyrkin, K. G. and Kniehl, Bernd A. and Steinhauser, M.",
    title = "{Strong Coupling Constant with Flavor Thresholds at Four Loops in the Modified Minimal-Subtraction Scheme}",
    eprint = "hep-ph/9706430",
    archivePrefix = "arXiv",
    reportNumber = "MPI-PHT-97-025",
    doi = "10.1103/PhysRevLett.79.2184",
    journal = "Phys. Rev. Lett.",
    volume = "79",
    pages = "2184--2187",
    year = "1997"
}

@article{Esteban:2020cvm,
    author = "Esteban, Ivan and Gonzalez-Garcia, M. C. and Maltoni, Michele and Schwetz, Thomas and Zhou, Albert",
    title = "{The fate of hints: updated global analysis of three-flavor neutrino oscillations}",
    eprint = "2007.14792",
    archivePrefix = "arXiv",
    primaryClass = "hep-ph",
    reportNumber = "IFT-UAM/CSIC-112, YITP-SB-2020-21",
    doi = "10.1007/JHEP09(2020)178",
    journal = "JHEP",
    volume = "09",
    pages = "178",
    year = "2020"
}

@article{Aoki:2017puj,
    author = "Aoki, Yasumichi and Izubuchi, Taku and Shintani, Eigo and Soni, Amarjit",
    title = "{Improved lattice computation of proton decay matrix elements}",
    eprint = "1705.01338",
    archivePrefix = "arXiv",
    primaryClass = "hep-lat",
    reportNumber = "RBRC-1235, KEK-CP-358",
    doi = "10.1103/PhysRevD.96.014506",
    journal = "Phys. Rev. D",
    volume = "96",
    number = "1",
    pages = "014506",
    year = "2017"
}

@article{Yoo:2021gql,
    author = "Yoo, Jun-Sik and Aoki, Yasumichi and Boyle, Peter and Izubuchi, Taku and Soni, Amarjit and Syritsyn, Sergey",
    title = "{Proton decay matrix elements on the lattice at physical pion mass}",
    eprint = "2111.01608",
    archivePrefix = "arXiv",
    primaryClass = "hep-lat",
    reportNumber = "RBRC-1333, KEK-CP-0385",
    doi = "10.1103/PhysRevD.105.074501",
    journal = "Phys. Rev. D",
    volume = "105",
    number = "7",
    pages = "074501",
    year = "2022"
}

@article{Super-Kamiokande:2020wjk,
    author = "Takenaka, A. and others",
    collaboration = "Super-Kamiokande",
    title = "{Search for proton decay via $p\to e^+\pi^0$ and $p\to \mu^+\pi^0$ with an enlarged fiducial volume in Super-Kamiokande I-IV}",
    eprint = "2010.16098",
    archivePrefix = "arXiv",
    primaryClass = "hep-ex",
    doi = "10.1103/PhysRevD.102.112011",
    journal = "Phys. Rev. D",
    volume = "102",
    number = "11",
    pages = "112011",
    year = "2020"
}

@article{Super-Kamiokande:2022egr,
    author = "Matsumoto, R. and others",
    collaboration = "Super-Kamiokande",
    title = "{Search for proton decay via $p\rightarrow \mu^+K^0$ in 0.37 megaton-years exposure of Super-Kamiokande}",
    eprint = "2208.13188",
    archivePrefix = "arXiv",
    primaryClass = "hep-ex",
    doi = "10.1103/PhysRevD.106.072003",
    journal = "Phys. Rev. D",
    volume = "106",
    number = "7",
    pages = "072003",
    year = "2022"
}

@article{Super-Kamiokande:2024qbv,
    author = "Taniuchi, N. and others",
    collaboration = "Super-Kamiokande",
    title = "{Search for proton decay via p{\textrightarrow}e+{\ensuremath{\eta}} and p{\textrightarrow}{\ensuremath{\mu}}+{\ensuremath{\eta}} with a 0.37~Mton-year exposure of Super-Kamiokande}",
    eprint = "2409.19633",
    archivePrefix = "arXiv",
    primaryClass = "hep-ex",
    doi = "10.1103/PhysRevD.110.112011",
    journal = "Phys. Rev. D",
    volume = "110",
    number = "11",
    pages = "112011",
    year = "2024"
}

@article{Hyper-Kamiokande:2018ofw,
    author = "Abe, K. and others",
    collaboration = "Hyper-Kamiokande",
    title = "{Hyper-Kamiokande Design Report}",
    eprint = "1805.04163",
    archivePrefix = "arXiv",
    primaryClass = "physics.ins-det",
    month = "5",
    year = "2018"
}

@article{Super-Kamiokande:2005lev,
    author = "Kobayashi, K. and others",
    collaboration = "Super-Kamiokande",
    title = "{Search for nucleon decay via modes favored by supersymmetric grand unification models in Super-Kamiokande-I}",
    eprint = "hep-ex/0502026",
    archivePrefix = "arXiv",
    doi = "10.1103/PhysRevD.72.052007",
    journal = "Phys. Rev. D",
    volume = "72",
    pages = "052007",
    year = "2005"
}

@article{Super-Kamiokande:2013rwg,
    author = "Abe, K. and others",
    collaboration = "Super-Kamiokande",
    title = "{Search for Nucleon Decay via $n \to \bar{\nu} \pi^{0}$ and $p \to \bar{\nu} \pi^{+}$ in Super-Kamiokande}",
    eprint = "1305.4391",
    archivePrefix = "arXiv",
    primaryClass = "hep-ex",
    doi = "10.1103/PhysRevLett.113.121802",
    journal = "Phys. Rev. Lett.",
    volume = "113",
    number = "12",
    pages = "121802",
    year = "2014"
}

@article{Super-Kamiokande:2014otb,
    author = "Abe, K. and others",
    collaboration = "Super-Kamiokande",
    title = "{Search for proton decay via $p\to\nu K^+$ using 260  kiloton\textperiodcentered{}year data of Super-Kamiokande}",
    eprint = "1408.1195",
    archivePrefix = "arXiv",
    primaryClass = "hep-ex",
    doi = "10.1103/PhysRevD.90.072005",
    journal = "Phys. Rev. D",
    volume = "90",
    number = "7",
    pages = "072005",
    year = "2014"
}

@article{JUNO:2015zny,
    author = "An, Fengpeng and others",
    collaboration = "JUNO",
    title = "{Neutrino Physics with JUNO}",
    eprint = "1507.05613",
    archivePrefix = "arXiv",
    primaryClass = "physics.ins-det",
    doi = "10.1088/0954-3899/43/3/030401",
    journal = "J. Phys. G",
    volume = "43",
    number = "3",
    pages = "030401",
    year = "2016"
}

@article{DUNE:2020fgq,
    author = "Abi, B. and others",
    collaboration = "DUNE",
    title = "{Prospects for beyond the Standard Model physics searches at the Deep Underground Neutrino Experiment}",
    eprint = "2008.12769",
    archivePrefix = "arXiv",
    primaryClass = "hep-ex",
    reportNumber = "FERMILAB-PUB-20-459-LBNF-ND",
    doi = "10.1140/epjc/s10052-021-09007-w",
    journal = "Eur. Phys. J. C",
    volume = "81",
    number = "4",
    pages = "322",
    year = "2021"
}

@article{Froggatt:1978nt,
    author = "Froggatt, C. D. and Nielsen, Holger Bech",
    title = "{Hierarchy of Quark Masses, Cabibbo Angles and CP Violation}",
    reportNumber = "CERN-TH-2519",
    doi = "10.1016/0550-3213(79)90316-X",
    journal = "Nucl. Phys. B",
    volume = "147",
    pages = "277--298",
    year = "1979"
}

@article{Pradler:2006hh,
    author = "Pradler, Josef and Steffen, Frank Daniel",
    title = "{Constraints on the Reheating Temperature in Gravitino Dark Matter Scenarios}",
    eprint = "hep-ph/0612291",
    archivePrefix = "arXiv",
    reportNumber = "MPP-2006-141",
    doi = "10.1016/j.physletb.2007.02.072",
    journal = "Phys. Lett. B",
    volume = "648",
    pages = "224--235",
    year = "2007"
}

@article{Cohen:2013ama,
    author = "Cohen, Timothy and Lisanti, Mariangela and Pierce, Aaron and Slatyer, Tracy R.",
    title = "{Wino Dark Matter Under Siege}",
    eprint = "1307.4082",
    archivePrefix = "arXiv",
    primaryClass = "hep-ph",
    reportNumber = "MIT-CTP-4482, SLAC-PUB-15664, MCTP-13-19",
    doi = "10.1088/1475-7516/2013/10/061",
    journal = "JCAP",
    volume = "10",
    pages = "061",
    year = "2013"
}

@article{Fan:2013faa,
    author = "Fan, JiJi and Reece, Matthew",
    title = "{In Wino Veritas? Indirect Searches Shed Light on Neutralino Dark Matter}",
    eprint = "1307.4400",
    archivePrefix = "arXiv",
    primaryClass = "hep-ph",
    doi = "10.1007/JHEP10(2013)124",
    journal = "JHEP",
    volume = "10",
    pages = "124",
    year = "2013"
}

@article{Ibe:2015tma,
    author = "Ibe, Masahiro and Matsumoto, Shigeki and Shirai, Satoshi and Yanagida, Tsutomu T.",
    title = "{Wino Dark Matter in light of the AMS-02 2015 Data}",
    eprint = "1504.05554",
    archivePrefix = "arXiv",
    primaryClass = "hep-ph",
    reportNumber = "DESY-15-054, IPMU-15-0053",
    doi = "10.1103/PhysRevD.91.111701",
    journal = "Phys. Rev. D",
    volume = "91",
    number = "11",
    pages = "111701",
    year = "2015"
}

@article{Kawasaki:2015yya,
    author = "Kawasaki, Masahiro and Kohri, Kazunori and Moroi, Takeo and Takaesu, Yoshitaro",
    title = "{Revisiting Big-Bang Nucleosynthesis Constraints on Dark-Matter Annihilation}",
    eprint = "1509.03665",
    archivePrefix = "arXiv",
    primaryClass = "hep-ph",
    reportNumber = "UT-15-34, IPMU-15-0160, KEK-TH-1863, KEK-COSMO-183",
    doi = "10.1016/j.physletb.2015.10.048",
    journal = "Phys. Lett. B",
    volume = "751",
    pages = "246--250",
    year = "2015"
}

@article{Slatyer:2015jla,
    author = "Slatyer, Tracy R.",
    title = "{Indirect dark matter signatures in the cosmic dark ages. I. Generalizing the bound on s-wave dark matter annihilation from Planck results}",
    eprint = "1506.03811",
    archivePrefix = "arXiv",
    primaryClass = "hep-ph",
    reportNumber = "MIT-CTP-4682",
    doi = "10.1103/PhysRevD.93.023527",
    journal = "Phys. Rev. D",
    volume = "93",
    number = "2",
    pages = "023527",
    year = "2016"
}

@article{Kawasaki:2021etm,
    author = "Kawasaki, Masahiro and Nakatsuka, Hiromasa and Nakayama, Kazunori and Sekiguchi, Toyokazu",
    title = "{Revisiting CMB constraints on dark matter annihilation}",
    eprint = "2105.08334",
    archivePrefix = "arXiv",
    primaryClass = "astro-ph.CO",
    doi = "10.1088/1475-7516/2021/12/015",
    journal = "JCAP",
    volume = "12",
    number = "12",
    pages = "015",
    year = "2021"
}

@article{Fermi-LAT:2015kyq,
    author = "Ackermann, M. and others",
    collaboration = "Fermi-LAT",
    title = "{Updated search for spectral lines from Galactic dark matter interactions with pass 8 data from the Fermi Large Area Telescope}",
    eprint = "1506.00013",
    archivePrefix = "arXiv",
    primaryClass = "astro-ph.HE",
    reportNumber = "FERMILAB-PUB-15-673-AE",
    doi = "10.1103/PhysRevD.91.122002",
    journal = "Phys. Rev. D",
    volume = "91",
    number = "12",
    pages = "122002",
    year = "2015"
}

@article{MAGIC:2022acl,
    author = "Abe, H. and others",
    collaboration = "MAGIC",
    title = "{Search for Gamma-Ray Spectral Lines from Dark Matter Annihilation up to 100~TeV toward the Galactic Center with MAGIC}",
    eprint = "2212.10527",
    archivePrefix = "arXiv",
    primaryClass = "astro-ph.HE",
    reportNumber = "KEK-TH-2487, KEK-Cosmo-0307",
    doi = "10.1103/PhysRevLett.130.061002",
    journal = "Phys. Rev. Lett.",
    volume = "130",
    number = "6",
    pages = "061002",
    year = "2023"
}

@article{Hisano:2003ec,
    author = "Hisano, Junji and Matsumoto, Shigeki and Nojiri, Mihoko M.",
    title = "{Explosive dark matter annihilation}",
    eprint = "hep-ph/0307216",
    archivePrefix = "arXiv",
    reportNumber = "ICRR-REPORT-500-2003-4, YITP-03-42",
    doi = "10.1103/PhysRevLett.92.031303",
    journal = "Phys. Rev. Lett.",
    volume = "92",
    pages = "031303",
    year = "2004"
}

@article{Hisano:2004ds,
    author = "Hisano, Junji and Matsumoto, Shigeki. and Nojiri, Mihoko M. and Saito, Osamu",
    title = "{Non-perturbative effect on dark matter annihilation and gamma ray signature from galactic center}",
    eprint = "hep-ph/0412403",
    archivePrefix = "arXiv",
    reportNumber = "ICRR-REPORT-513-2004-11, YITP-04-73",
    doi = "10.1103/PhysRevD.71.063528",
    journal = "Phys. Rev. D",
    volume = "71",
    pages = "063528",
    year = "2005"
}

\end{document}